\RequirePackage{lineno}
\documentclass[twocolumn,showpacs,superscriptaddress,amsmath,amssymb,nofootinbib]{revtex4-1}
\usepackage{graphicx}
\usepackage{epsfig}
\usepackage{overpic}
\usepackage{dcolumn}
\usepackage{ulem}
\usepackage{bm}
\usepackage{lineno}
\usepackage{xspace}
\usepackage{multirow}
\usepackage{epstopdf}
\usepackage{xcolor}
\usepackage{soul}
\usepackage[colorlinks,allcolors=black]{hyperref}
\usepackage{verbatim}
\usepackage{appendix}
\usepackage{makecell}

\usepackage{subfigure}
\include{def-com}

\newcommand{\gev}{\ensuremath{\;\text{GeV}\xspace}}

\newcommand{\nbar}{\bar{n}}

\begin{document}
\title{Measurements of the electric and magnetic form factors of the neutron for time-like momentum transfer}
%\linenumbers

\author{
\noindent\large{
%{\bf Author List:}
%\author{Author list}
\begin{small}
\begin{center}
%% Saved at => 2022-10-11
M.~Ablikim$^{1}$, M.~N.~Achasov$^{13,b}$, P.~Adlarson$^{73}$, R.~Aliberti$^{34}$, A.~Amoroso$^{72A,72C}$, M.~R.~An$^{38}$, Q.~An$^{69,56}$, Y.~Bai$^{55}$, O.~Bakina$^{35}$, I.~Balossino$^{29A}$
% ...
\\
\vspace{0.2cm}
(BESIII Collaboration)\\
\vspace{0.2cm} {\it
$^{1}$ Institute of High Energy Physics, Beijing 100049, People's Republic of China\\
% ...
$^{13}$ G.I. Budker Institute of Nuclear Physics SB RAS (BINP), Novosibirsk 630090, Russia\\
% ...
$^{29}$ INFN Sezione di Ferrara, (A)INFN Sezione di Ferrara, I-44122, Ferrara, Italy; (B)University of Ferrara,  I-44122, Ferrara, Italy\\
% ...
$^{34}$ Johannes Gutenberg University of Mainz, Johann-Joachim-Becher-Weg 45, D-55099 Mainz, Germany\\
$^{35}$ Joint Institute for Nuclear Research, 141980 Dubna, Moscow region, Russia\\
% ...
$^{38}$ Liaoning Normal University, Dalian 116029, People's Republic of China\\
% ...
$^{55}$ Southeast University, Nanjing 211100, People's Republic of China\\
$^{56}$ State Key Laboratory of Particle Detection and Electronics, Beijing 100049, Hefei 230026, People's Republic of China\\
% ...
$^{69}$ University of Science and Technology of China, Hefei 230026, People's Republic of China\\
% ...
$^{72}$ University of Turin and INFN, (A)University of Turin, I-10125, Turin, Italy; (B)University of Eastern Piedmont, I-15121, Alessandria, Italy; (C)INFN, I-10125, Turin, Italy\\
$^{73}$ Uppsala University, Box 516, SE-75120 Uppsala, Sweden\\
% ...
$^{b}$ Also at the Novosibirsk State University, Novosibirsk, 630090, Russia\\
% ...
}
\end{center}
\vspace{0.4cm}
\end{small}
}}

%% ends here %%

%\date{\today}
\begin{abstract}
We present the first measurements of the electric and magnetic form factors of the neutron in the time-like (positive $q^2$) region as function of four-momentum transfer.
We explored the differential cross sections of the reaction $e^+e^- \rightarrow \bar{n}n$ with data collected with the BESIII detector at the BEPCII accelerator, corresponding to an integrated luminosity of 354.6~pb$^{-1}$ in total at twelve center-of-mass energies between $\sqrt{s} = 2.0 - 2.95$ GeV.
A relative uncertainty of 18\% and 12\% for the electric and magnetic form factors, respectively, is achieved at $\sqrt{s} = 2.3935$ GeV.
Our results are comparable in accuracy to those from electron scattering in the comparable space-like (negative $q^2$) region of four-momentum transfer.
The electromagnetic form factor ratio  $R_{\rm em}\equiv |G_E|/|G_M|$ is within the uncertainties close to unity. We compare our result on $|G_E|$ and $|G_M|$
to recent model predictions, and the measurements in the space-like region to test the analyticity of electromagnetic form factors.
\end{abstract}
\maketitle

\begin{figure}[h]
        \centering
        \begin{overpic}[height=3.3cm,width=3.2cm]{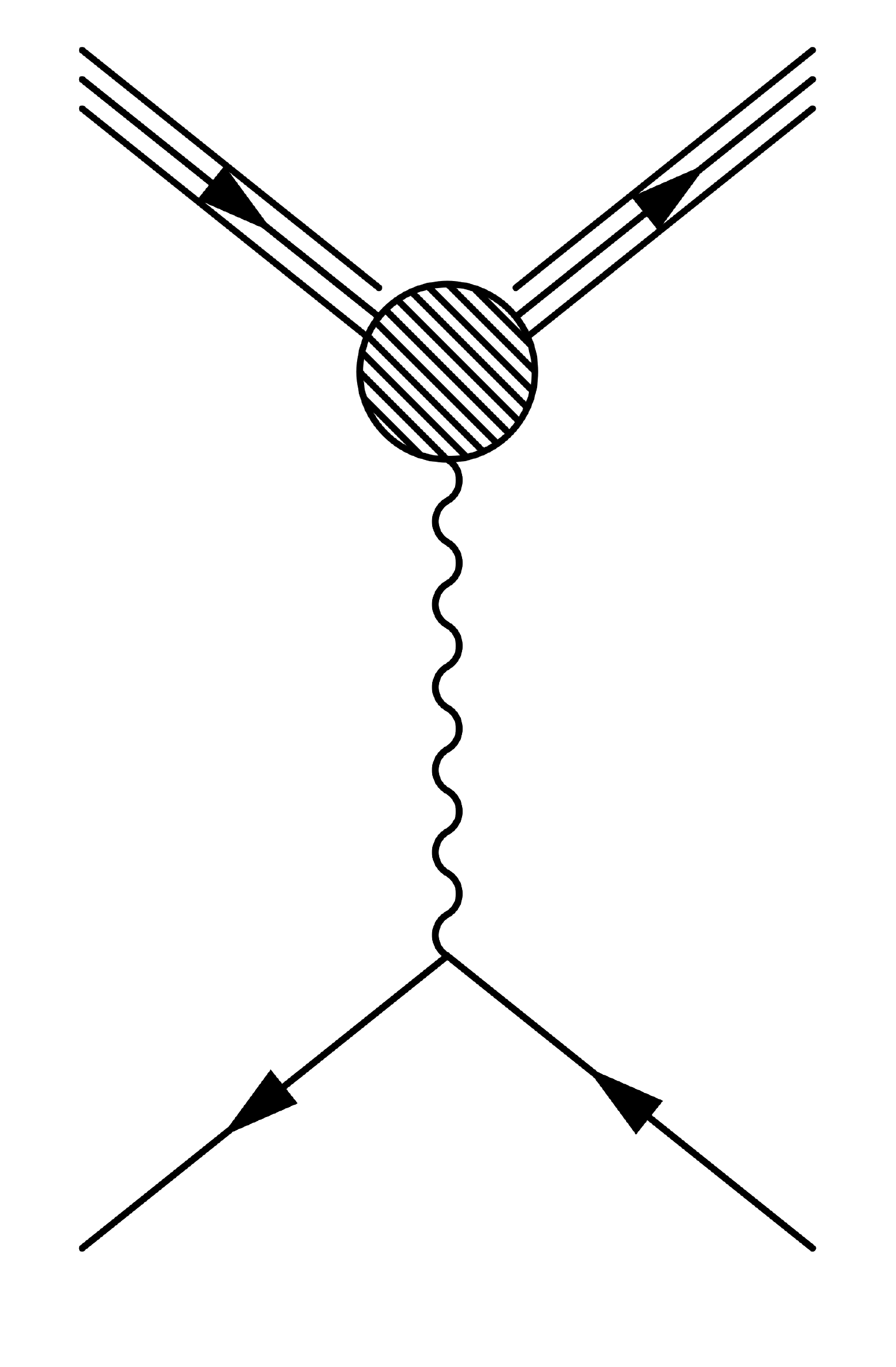}
           \put(20, 0){\Large{{\rm${\bf e^-}$}}}
            \put(66, 0){\Large{{\rm${\bf e^-}$}}}
            \put(20, 92){\Large{{\rm${\bf n}$}}}
            \put(66, 92){\Large{{\rm${\bf n}$}}}
            \put(28, 42){\large{{\rm${\boldsymbol \gamma^{*}}$}}}
            \put(52, 42){\large{{\rm${\bf q^2 < 0}$}}}
        \end{overpic}
        \hspace{0.2cm}
        \begin{overpic}[height=3.2cm, width=5.cm]{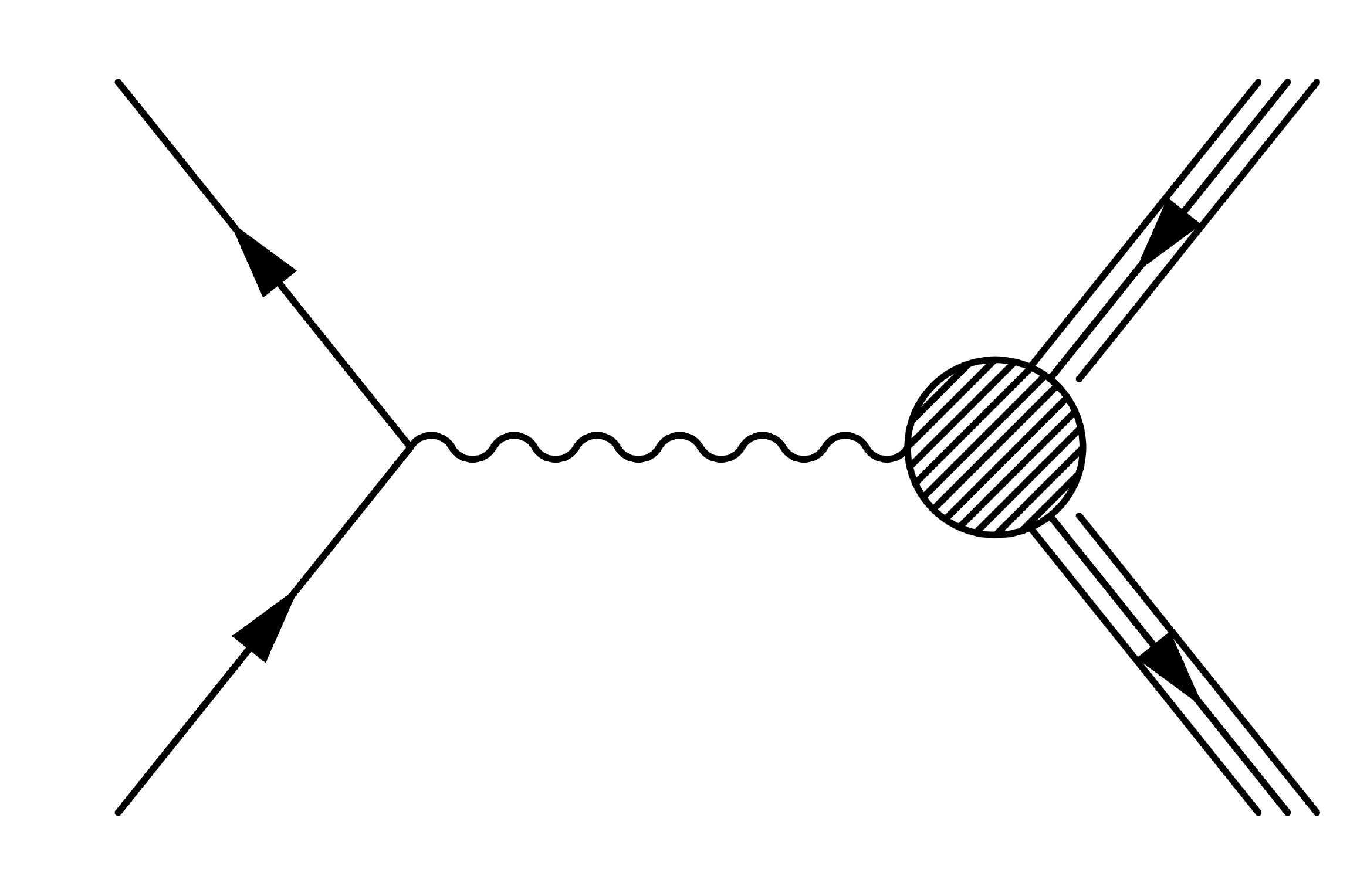}
            \put(8, 60){\Large{{\rm${\bf e^+}$}}}
            \put(8, 0){\Large{{\rm${\bf e^-}$}}}
            \put(85, 60){\Large{{\rm${\bf \bar{n}}$}}}
            \put(85, 0){\Large{{\rm${\bf n}$}}}
            \put(45, 38){\large{{\rm${\boldsymbol \gamma^{*}}$}}}
            \put(30, 22){\large{{\rm${\bf q^2 > 4M_n^2}$}}}
        \end{overpic}
\caption{The lowest order Feynman diagram for the SL process (left) $e^- n\to e^- n$ and the TL process (right) $e^+e^-\to n \bar{n}$.
Here, $\gamma^*$ represents the virtual photon transferring the four-momentum squared $q^2$ of the reaction. The gray circle represents the internal nucleon structure parameterized by the EMFFs. }
\label{0}
\end{figure}
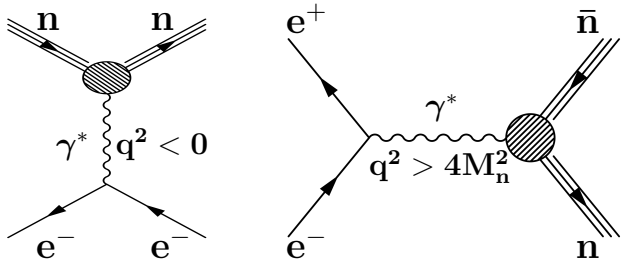

Proton and neutron are the fundamental building blocks of atomic nuclei.
Their complex internal structure emerges from Quantum Chromodynamics (QCD) but is not accessible from ab initio calculations
in the non-perturbative regime of QCD governed by quark confinement~\cite{Wilson:1974sk} and gluon self-coupling~\cite{gsc}.
On the other hand, measurements of the electromagnetic form factors (EMFFs) of the nucleon are straightforward~\cite{Hofstadter:1955ae}.
EMFFs have long since served as a testing ground for the understanding of QCD at low momentum transfer $q^2$. Since Hofstadter's groundbreaking measurements~\cite{Hofstadter:1956Review}, various experiments at different facilities successfully measured nucleon EMFFs~\cite{Punjabi:2015bba} in electron scattering with increasing precision, providing important input for theoretical calculations of nucleon properties~\cite{mass,spin}, in particular the size of the neutron charge radius~\cite{Atac:2021wqj}.
A spin-$\frac{1}{2}$ particle, such as the nucleon, is described by two EMFFs, $G_E(q^2)$ and $G_M(q^2)$,
 which are Fourier-transforms of the intrinsic electric and magnetic distributions of the nucleon in the Breit frame~\cite{sachs}.
Depending on the sign of $q^{2}$ of the virtual exchange photon, we can distinguish two types of reactions.
The space-like (SL) region of negative $q^{2}$ can be accessed in lepton scattering, the time-like (TL) region of positive $q^2$ in annihilation reactions (see Fig.~\ref{0}).

Precise measurements of scattering of leptons with neutrons  \cite{reviewSL1,reviewSL2,reviewSL3,Yearian:1958shh} are far more difficult than with protons  due to the absence of free neutron targets. Possible alternatives like deuterons inevitably introduce uncertainties from nuclear binding corrections. On the other hand, free protons and neutrons are directly accessible in the TL region.
Measurements of the TL EMFFs of the proton have significantly gained precision over recent years~\cite{exp9,BaBar:2013ves, bes3ppbar, bes3ppbar2, bes3ppbar3, bes3ppbar4}. However, data with large statistics from $e^+e^-$ annihilation reactions has been rare. The first measurements of the neutron TL EMFFs were reported in the 1990s by the FENICE~\cite{FENICE} and the DM2~\cite{DM2} experiments and  two other measurements were reported by the SND experiment~\cite{Achasov:2014ncd,Achasov:2022ryd}.
A precise measurement of the neutron effective form factors $|G|$  was recently published by the BESIII collaboration~\cite{np}. However, so far no separate result for $G_E$ and $G_M$ is  available due to the difficulties in (anti-)neutron detection and efficiency calibration. These difficulties have prevented a detailed analysis of angular-dependent differential cross sections.

In this letter
we report the first measurement of separate neutron EMFF moduli $|G_E|$ and $|G_M|$ in the TL regime by exploring the differential cross sections of the reaction $e^+e^- \rightarrow \bar{n}n$
with the data collected by the BESIII~\cite{bes3detector} experiment at the BEPCII~\cite{bepc2} collider.
Compared with the total cross sections, the differential cross sections provide more information to determine scattering amplitudes of distinguished partial waves~\cite{Yang:2022qoy} in order to verify the analyticity of EMFFs and to test various nucleon models.
The data analyzed in this work corresponds to a total integrated luminosity $\mathcal{L}_{int}=354.6$ pb$^{-1}$~\cite{bes3:lum1,bes3:lum2}
at twelve center-of-mass energies (c.m.) between $\sqrt{s}=2.0$ and 2.95 GeV, and is grouped into five energy intervals to extract the EMFFs. An additional dataset with $10087\pm44$  million $J/\psi$ events~\cite{totaljpsi} has been used for a precise data-driven calibration of the $n (\bar{n})$ detection efficiency with the processes $J/\psi\to \bar{p}\pi n(p \pi \bar{n})$, and an investigation of identification and reconstruction of neutral particles  using the processes $J/\psi\to  n\bar{n}$,  $J/\psi\to \pi^+\pi^-\pi^0 (\to \gamma\gamma)$ and $e^+e^-\to\gamma\gamma$.

The moduli of the electric and magnetic form factors of the neutron can be %simultaneously
obtained by comparing the theoretical prediction  to
the experimentally accessible differential Born cross section as discussed in~\cite{diffborn}:

\begin{eqnarray}\footnotesize
\centering
\begin{split}
&\frac{N^{bin}}{\mathcal{L}\mathcal{E}^{MC}\mathcal{E}^{cor}(1+\delta)}  =   \int_{bin} \frac{{\pi}{\alpha^2}{\beta}C}{2s} |G_M|^2 \Big[  (1 + \cos^2\theta_{\bar{n}})   &  \\
  & \qquad \qquad \qquad +\quad \tau |R_{em}|^2 \sin^2\theta_{\bar{n}}\Big] d\cos\theta_{\bar{n}}, \quad \tau= \frac{4M_n^2}{s}.
 \end{split}
\label{fitFunc2}
\end{eqnarray}	
Here, $s\equiv q^2$ represents the c.m. energy squared of the electron-positron system, $M_n$ the neutron mass,
$\beta=\sqrt{1-\tau}$ its velocity, and
$C$ the Coulomb enhancement factor accounting for the electromagnetic interaction between the outgoing baryons, which is equal to 1 for neutral baryons. Furthermore,
$\cos\theta_{\bar{n}}$ is the cosine of the $\bar{n}$ polar angle along the positron beam direction in the electron-positron c.m. frame,
$|G_M|$ the modulus of the magnetic form factor, and $R_{em}=|G_E|/|G_M|$ the ratio of the moduli of the electric and magnetic form factors.
The differential Born cross section Eq.~(\ref{fitFunc2}), is calculated using the luminosity $\mathcal{L}$,
the signal reconstruction efficiency $\mathcal{E}^{MC}$ and
its correction $\mathcal{E}^{cor}$,
the next-to-leading-order radiation and vacuum-polarization
correction (1+$\delta$),
and the signal yield $N^{bin}$
per $\cos\theta_{\bar{n}}$ bin.
We integrate over the bin width in accordance to histograms of angular distribution as will be discussed below.

The final state of the signal process contains
 one antineutron and one neutron. Hence our data analysis strategy is based on the rejection of
events with charged tracks in the multilayer drift chamber (MDC).
For each event, the most energetic shower in the electromagnetic calorimeter (EMC) within the polar angle range of $|\cos\theta | < 0.7$ and a deposited energy within (0.5, 2.0)~GeV is considered as a $\nbar$ candidate. This shower is further associated with a response in the time-of-flight~(TOF) system aggregating all hits within an azimuthal angle span of 6 TOF plastic scintillators along the $\bar{n}$ momentum. To avoid a potential bias and to provide a cross-check, events are classified into three categories (i = A,B,C) depending on the detector responses to  $n\bar{n}$ particles. Events with responses from knockoff protons interaction in the TOF plastic scintillators from both particles and one associated hadronic shower registered in the EMC from the antineutron are classified as category A. Events with showers in the EMC from both particles, but only one measured knockoff proton interaction in the TOF from the anti-neutron
are assigned to category B. Events lacking any TOF responses but with reconstructed hadronic showers in the EMC from both particles are classified as category C. Each event belongs to not more than one category.
These categories not only provide a reliable cross-check but also guarantee a better precision using an inverse-variance weighting technique.
The selection method for the three categories is described in details
in Ref.~\cite{np}.

\begin{figure}[h]
        \centering
        \begin{overpic}[scale=0.18]{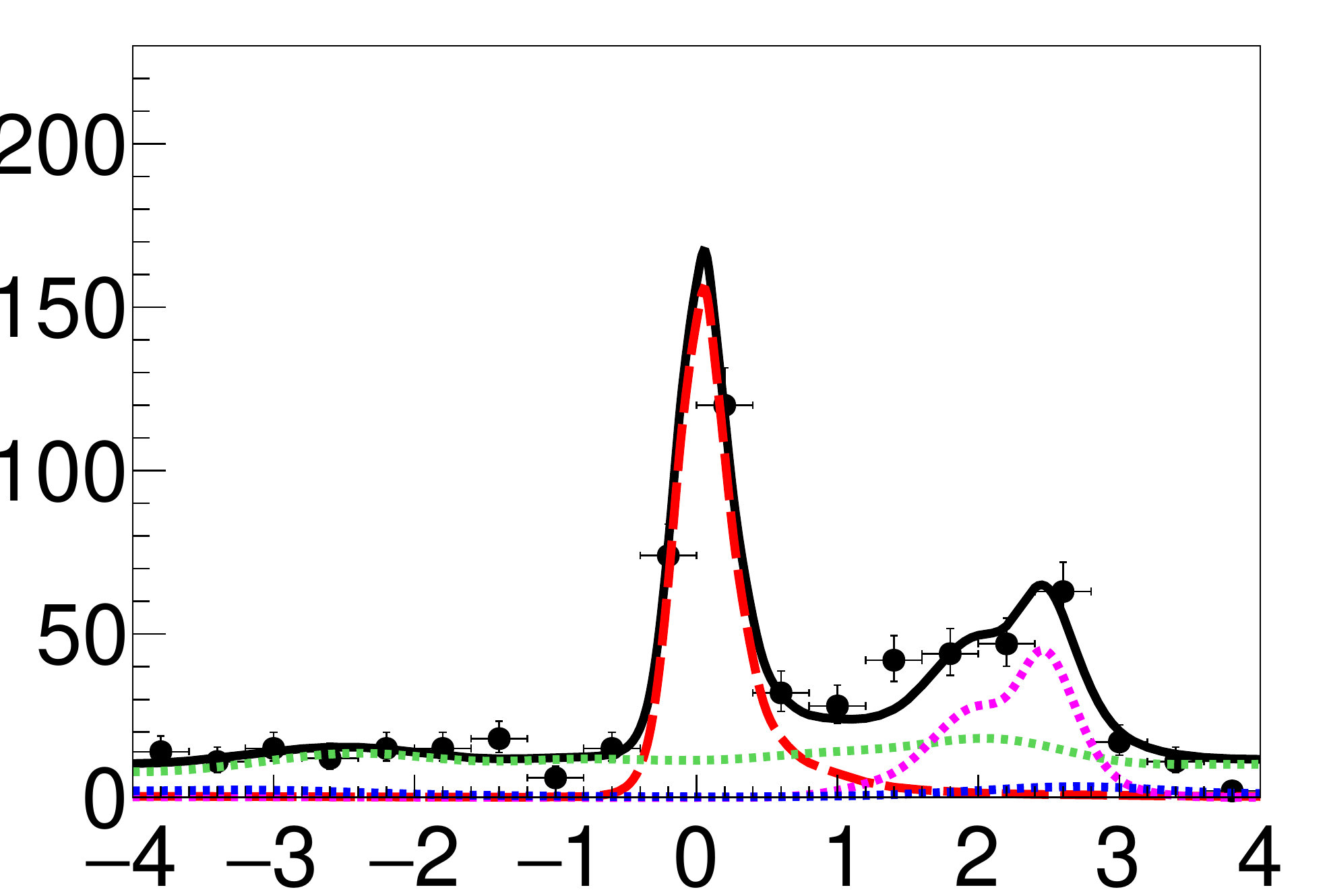}
            \put(35, 55){\scriptsize{{\rm${\bf |\bf\cos\theta_{\bar{n}}|<0.7}$}}}
        \end{overpic}
        \begin{overpic}[scale=0.18]{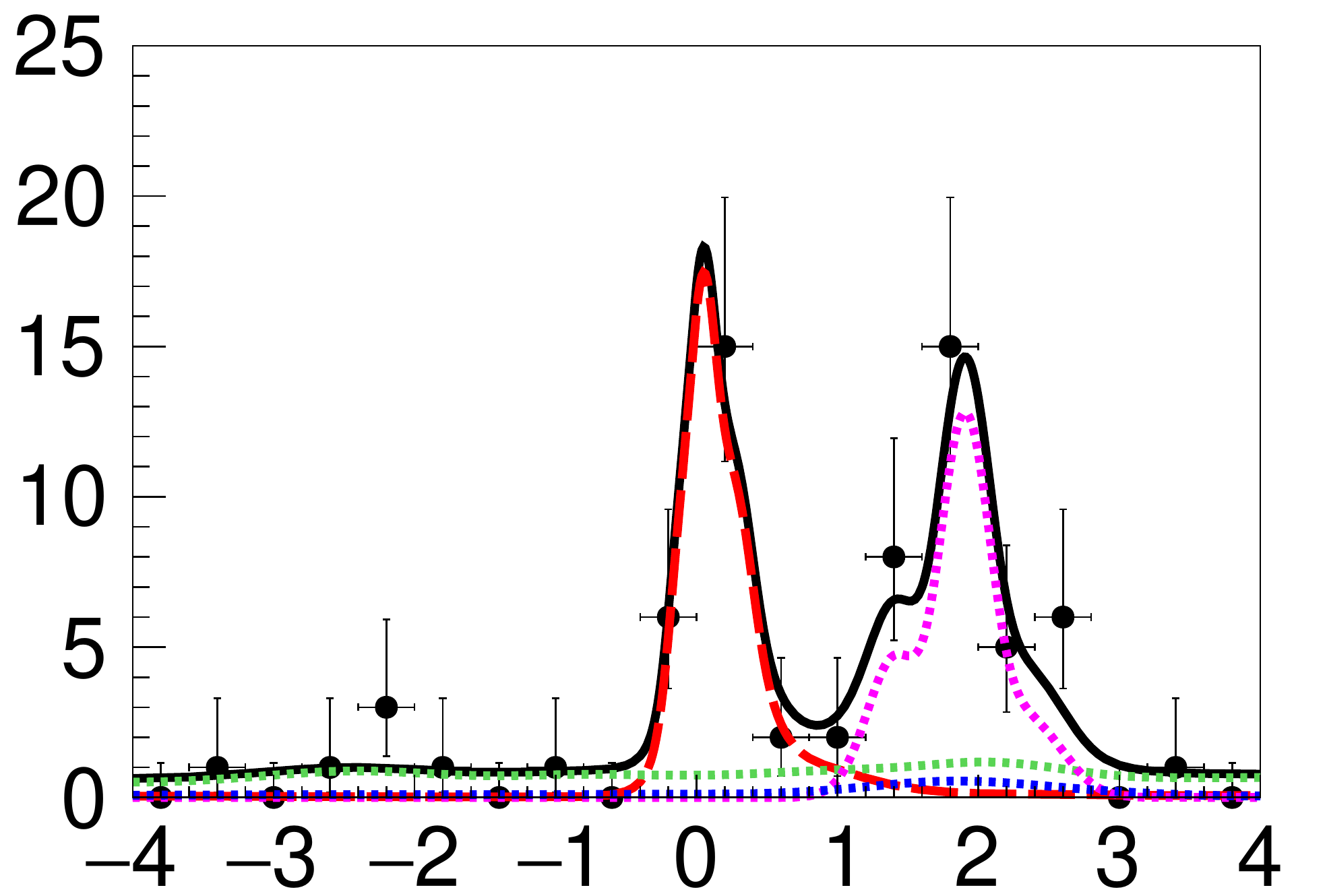}
            \put(20, 55){\scriptsize{{\rm${\bf -0.7<\cos\theta_{\bar{n}}<-0.5}$}}}
        \end{overpic}
        \begin{overpic}[scale=0.18]{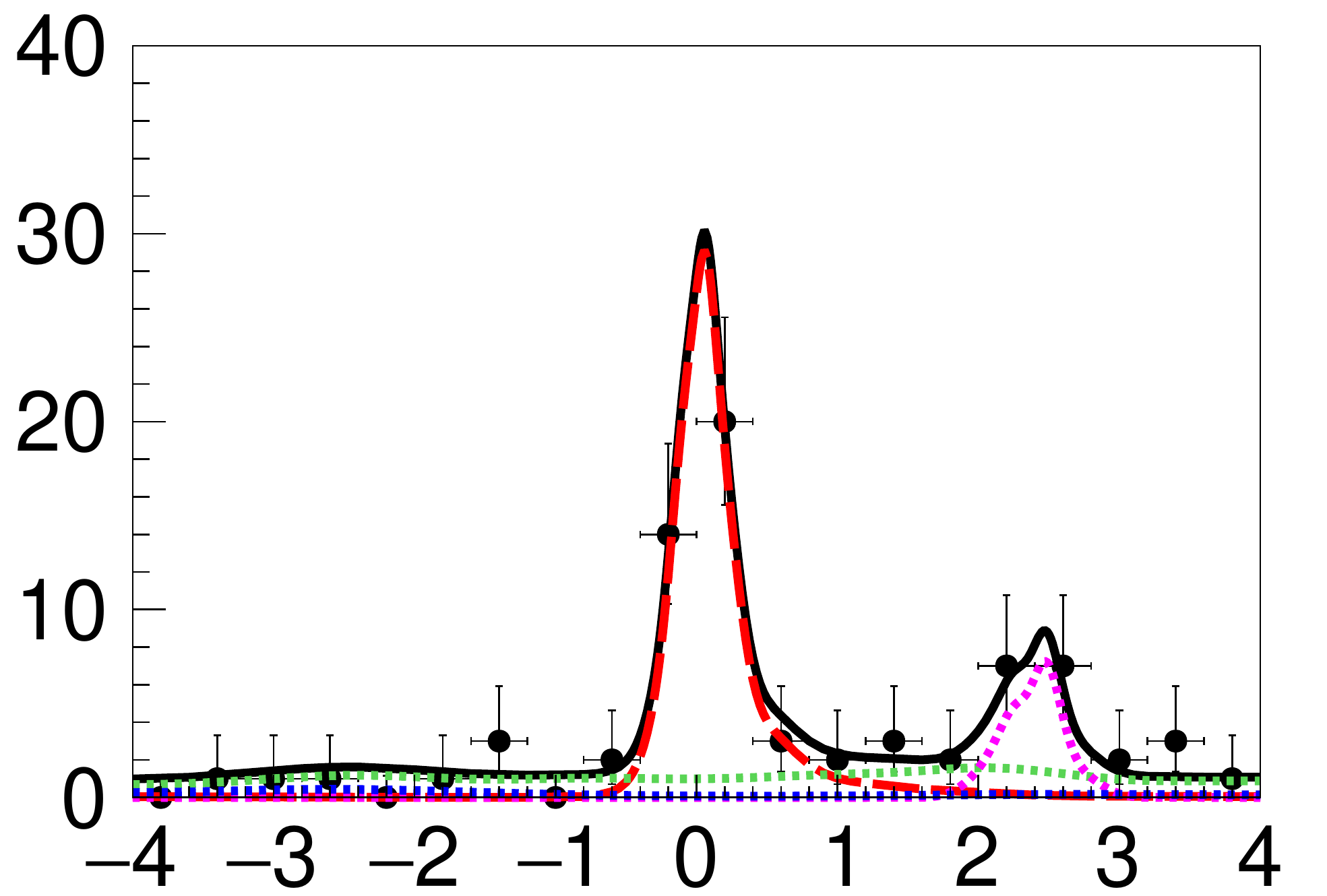}
            \put(20, 55){\scriptsize{{\rm${\bf -0.5<\cos\theta_{\bar{n}}<-0.3}$}}}
        \end{overpic}
        \begin{overpic}[scale=0.18]{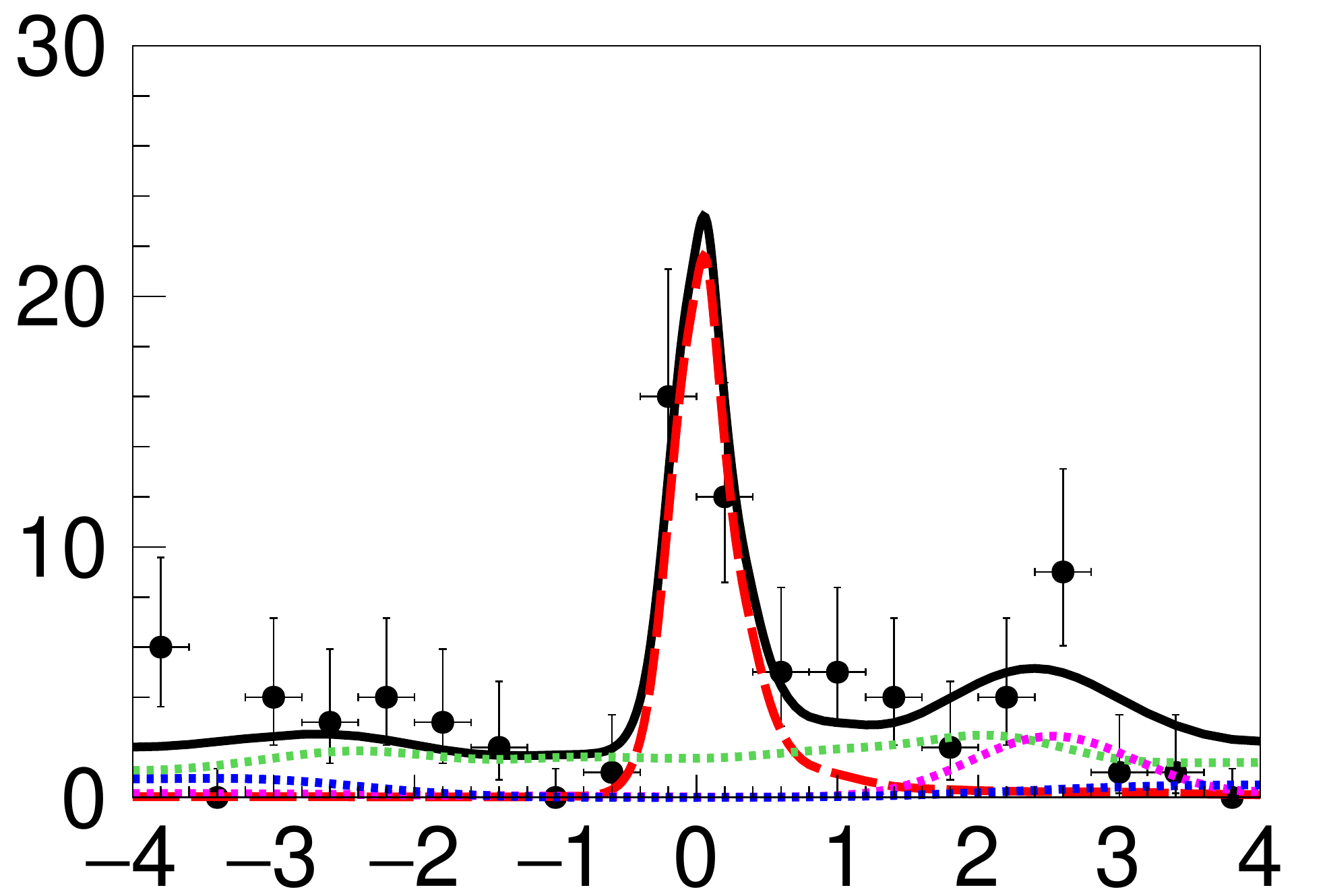}
            \put(20, 55){\scriptsize{{\rm${\bf -0.3<\cos\theta_{\bar{n}}<-0.1}$}}}
        \end{overpic}
        \begin{overpic}[scale=0.18]{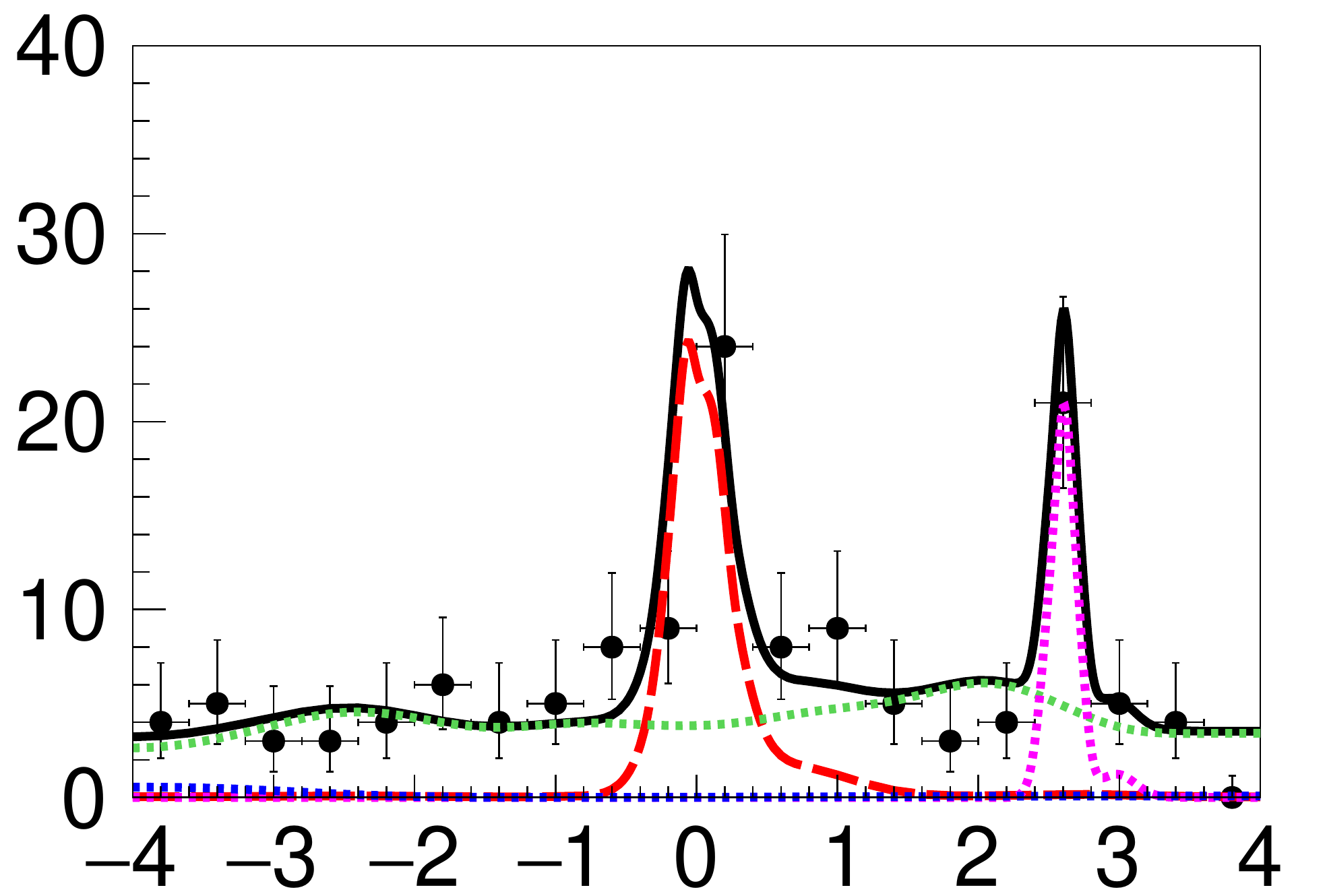}
            \put(23, 55){\scriptsize{{\rm${\bf -0.1<\cos\theta_{\bar{n}}<0.1}$}}}
            \put(-15, 30){\Large{{\rotatebox{90}{\bf  Events / 0.2}}}}
        \end{overpic}
        \begin{overpic}[scale=0.18]{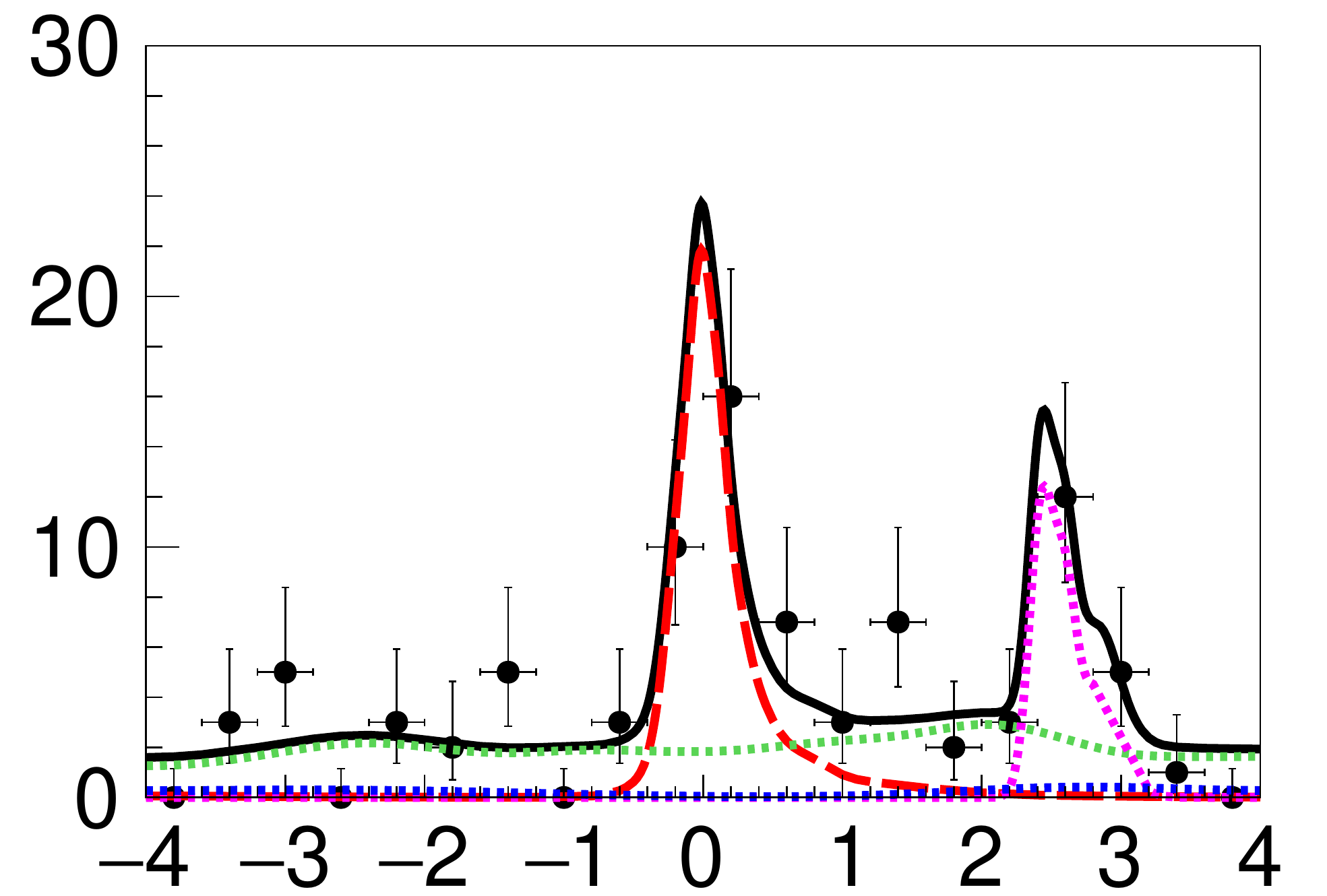}
            \put(28, 55){\scriptsize{{\rm${\bf 0.1<\cos\theta_{\bar{n}}<0.3}$}}}
        \end{overpic}
        \begin{overpic}[scale=0.18]{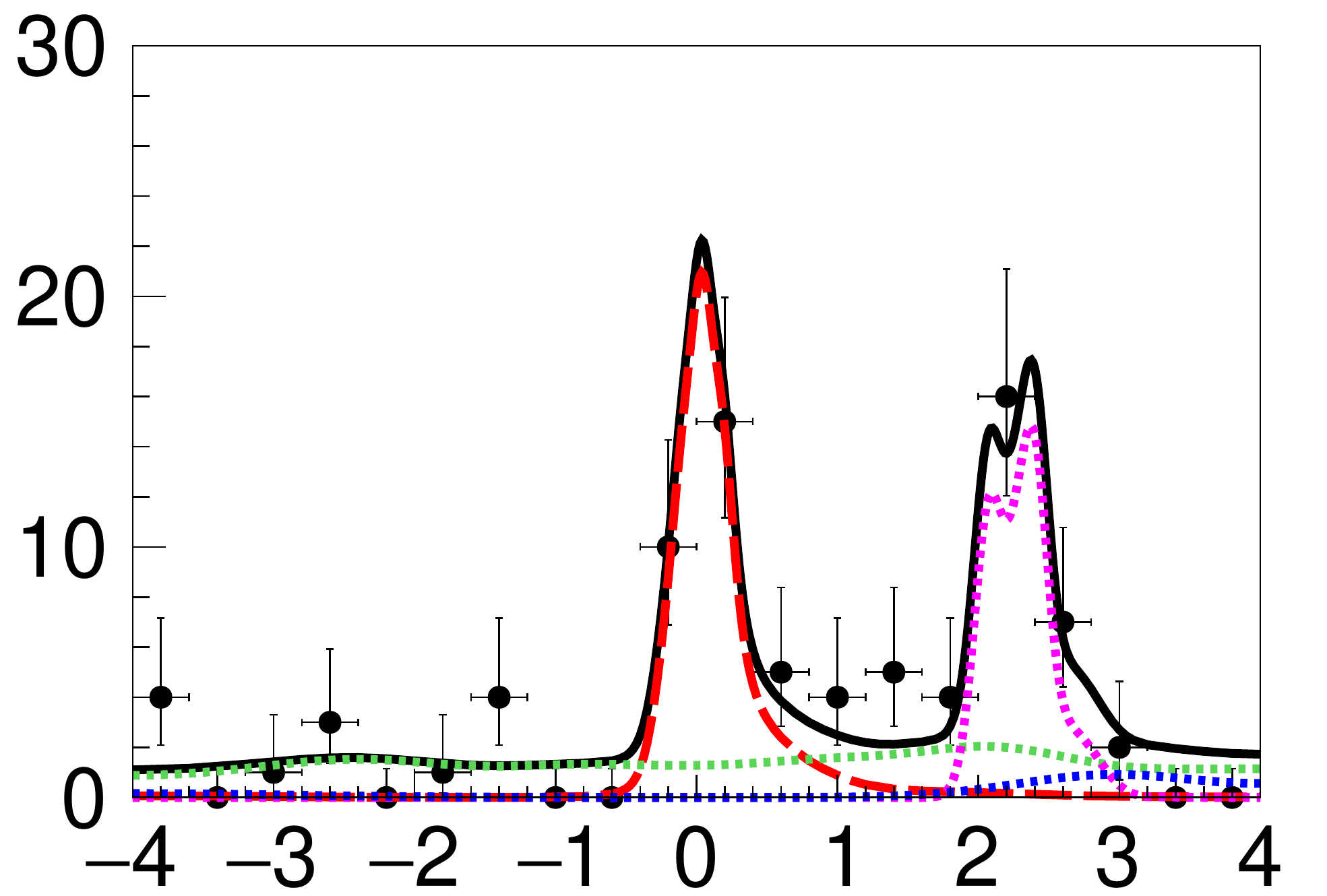}
            \put(28, 55){\scriptsize{{\rm${\bf 0.3<\cos\theta_{\bar{n}}<0.5}$}}}
                        \put(50, -10){\large{${\bf \Delta T_n = T_{TOF} - T_{exp} (ns)}$}}
        \end{overpic}
        \begin{overpic}[scale=0.18]{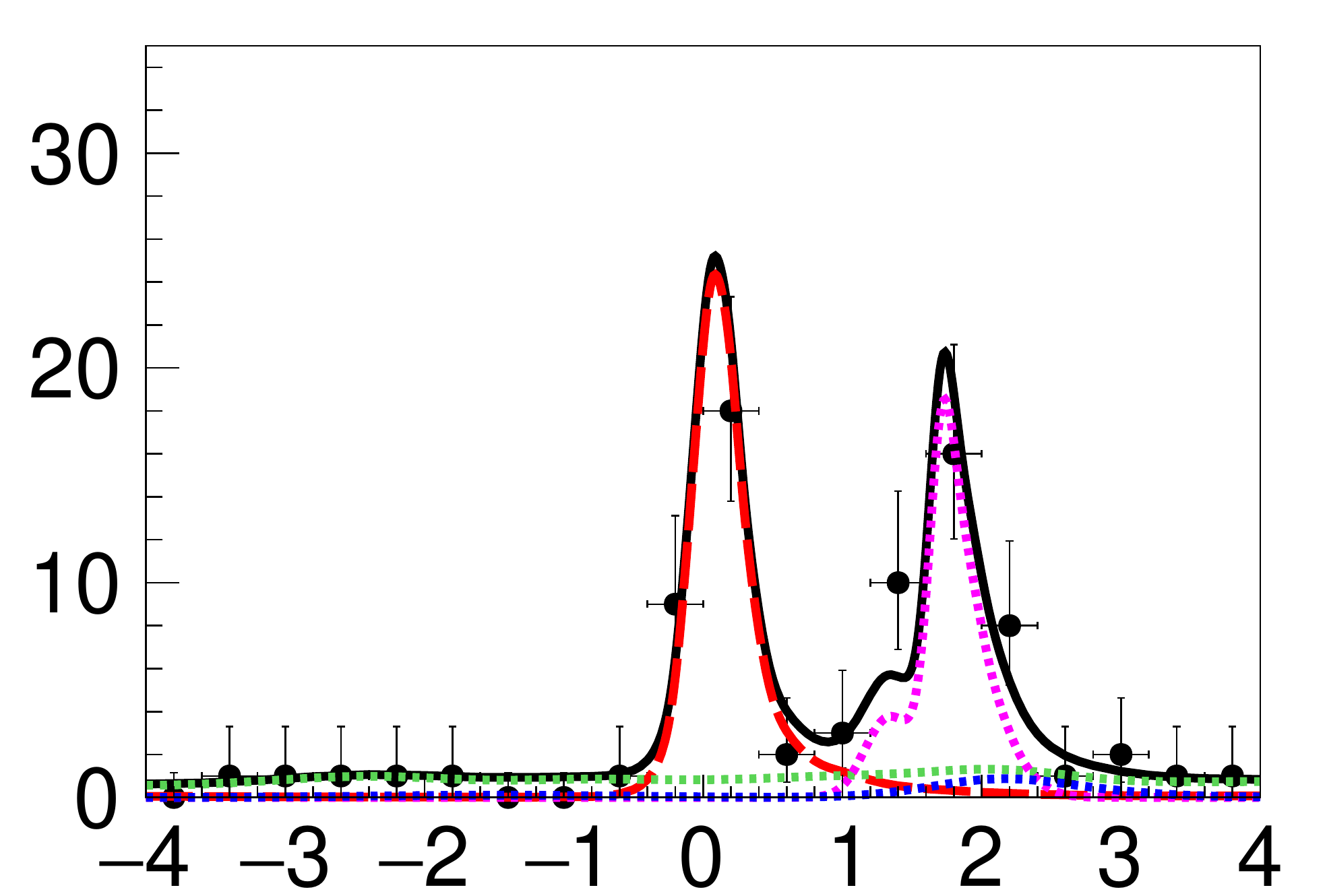}
            \put(28, 55){\scriptsize{{\rm${\bf 0.5<\cos\theta_{\bar{n}}<0.7}$}}}
        \end{overpic}\\
        \vspace{0.3cm}\caption{Fit to the $\Delta T_n$ distribution at $\sqrt{s}=2.1250$ GeV in 7 $\cos\theta_{\bar{n}}$ bins. Data are shown as black dots with error bars, the total fit as the black line, the signal component as the red dashed line, the $e^-e^+\to \gamma\gamma$ background components as the magenta dashed line, and the beam-related background components as the blue dashed line. The asymmetrical uncertainties for the data are determined from the Poisson distribution to take into account the low statistics in most bins.  }
        \label{supp1}
\end{figure}

With surviving events passing aforementioned selections per category,
$\cos\theta_{\bar{n}}$ of the antineutron is filled in histograms of 7 equidistant bins within $-0.7 < \cos\theta_{\bar{n}} < 0.7$.
For category A, signal events are characterized by the time difference between the time measured with the TOF and the expected flight time calculated from the neutron's momentum and flight path, respectively. For category B and C, signal events are characterized by the opening angle $\sphericalangle^n_{\bar{n}}$ between neutron and antineutron as measured with the EMC. Since the surviving events still contain contributions coming mainly from the beam-related background and the $e^+e^-\to\gamma\gamma$ background, we use a composite model taking into account the background ($\mathcal{M}^b$) and signal ($\mathcal{M}^s$) distributions to fit data and determine the number of reconstructed signal events $N^{bin}_i(\cos\theta_{\bar{n}})$  (i = A,B,C) at a given $\cos\theta_{\bar{n}}$ bin:
\begin{eqnarray}\footnotesize
\centering
\begin{split}
&\mathcal{M}_i(x) =  N^{bin}_i \mathcal{M}^s_i(x) + N^b_i \mathcal{M}^b_i(x), \\ & x=\Delta T_n\ \text{or}\ \sphericalangle^n_{\bar{n}},  \ \forall \cos\theta_{\bar{n}} \in (-0.7,0.7).
 \end{split}
\label{fitFunc3}
\end{eqnarray}
The signal (background) distribution are modeled with the MC simulation samples of the signal (background) process generated with {\sc CONEXC} event generator or control samples.
The fit optimization is performed for each category by minimizing a global negative log-likelihood (NLL) with the {\sc MIGRAD} \cite{roofit} routine, taking into account local NLLs from each data set (i=A,B,C) and using a modified version of the David-Fletcher-Powell method \cite{fletcher}.
Following, a {\sc HESSE} and a {\sc MINOS} algorithm \cite{roofit} improve the uncertainty determination of the model parameters, which describe the yields of both signal and background.
Note that, background distributions are shared among all energies for the same category. The asymmetrical uncertainties associated  to $N^{bin}_i$ are from the Poisson distribution to take into account the low statistics at most bins. As an example, Fig.~\ref{supp1} shows the extraction of signal yield $N^{bin}_{i=A}$ at $\sqrt{s} = 2.1250$ GeV for category A.

The signal reconstruction efficiency $\mathcal{E}_i(\cos\theta)$ for each category is determined by a dedicated Monte Carlo (MC) simulation for the process $e^{+}e^{-}\to n\bar{n}$ at each $\sqrt{s}$ using the MC generator {\sc Conexc}~\cite{ping} up to the next-to-leading order, followed by a {\sc Geant4}~\cite{geant4}-based simulation procedure, which mimics the response of various particles in the BESIII detector.
A data-driven efficiency calibration $\mathcal{C}_i^{dm}(\cos\theta)$ is achieved for both $n$ and $\bar{n}$ by using the process $J/\psi\to \bar{p}\pi n(p \pi \bar{n})$.
A trigger correction $\mathcal{C}_i^{trg}(\cos\theta)$ describing the probability of EMC-based online trigger capturing neutral processes is also applied. In addition, $(1+\delta)_i(\cos\theta)$ is the initial state radiation and vacuum polarization correction.
All corrections are multiplied to be $\mathcal{E}_{i}^{cor}(\cos\theta)=\mathcal{C}_i^{dm}\cdot \mathcal{C}_i^{trg}$.
The details about these corrections are given in Ref.~\cite{np}.

\begin{figure}[h]
        \centering
        \begin{overpic}[scale=0.20]{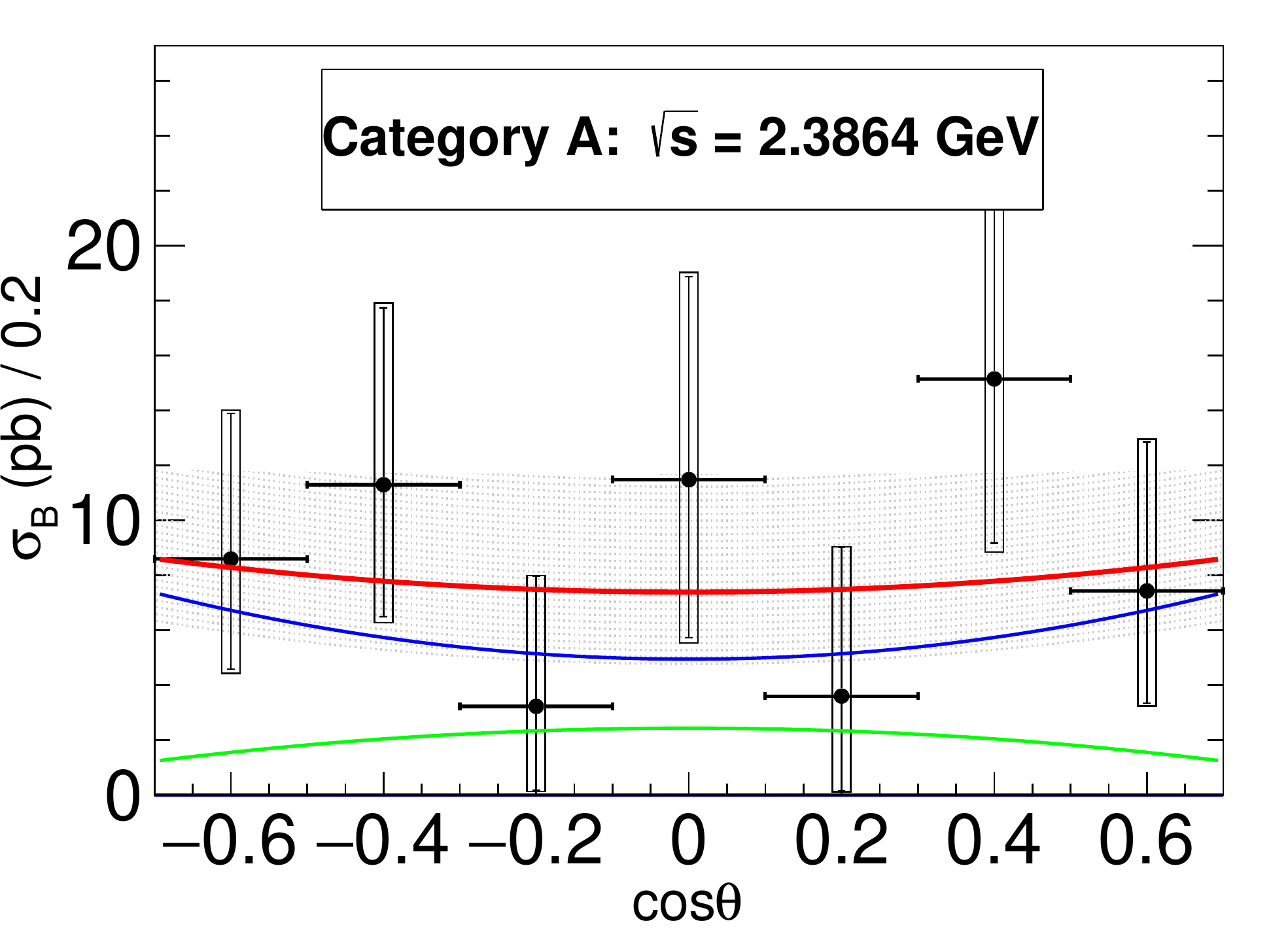}
        \end{overpic}
        \begin{overpic}[scale=0.20]{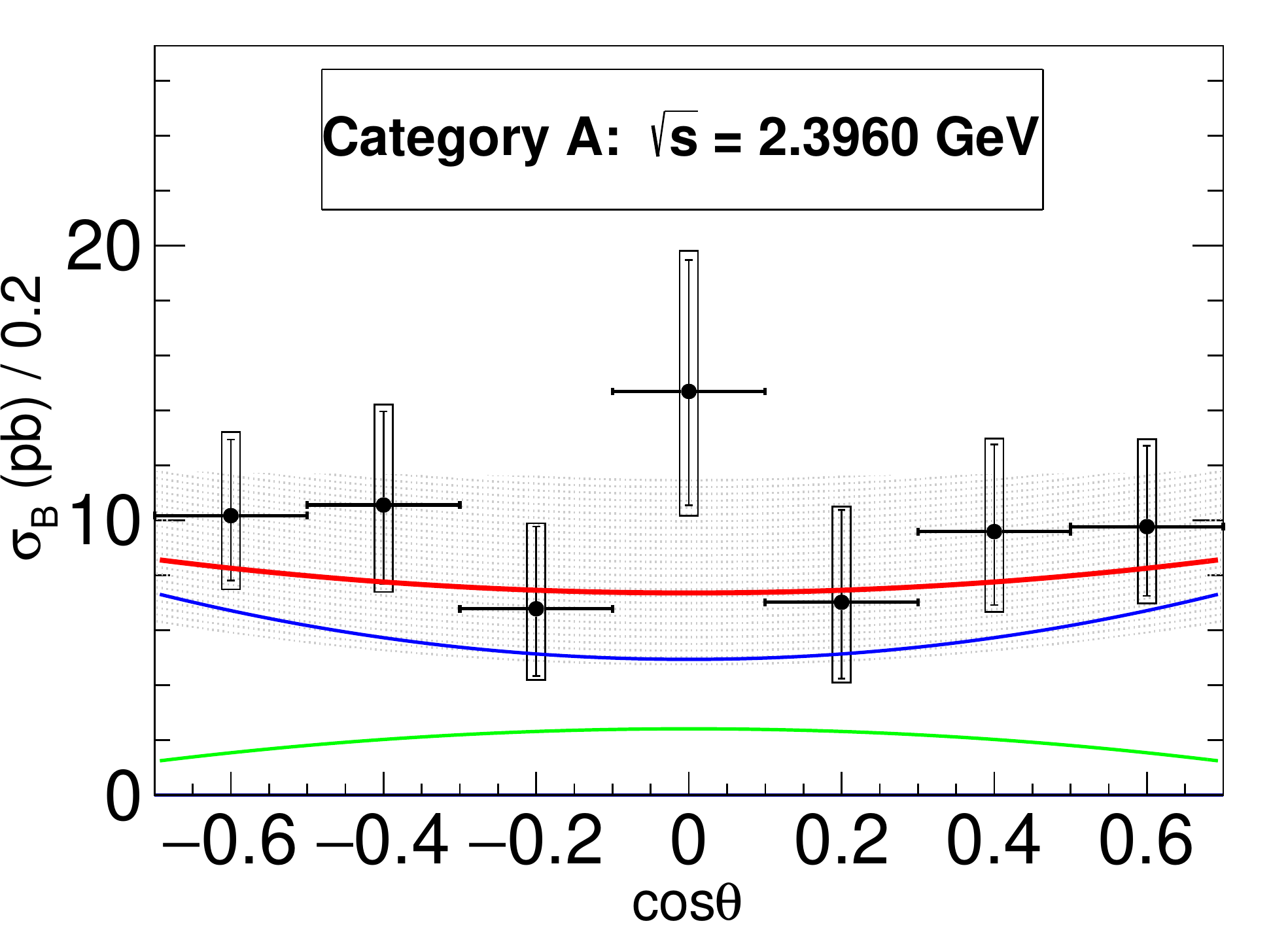}
        \end{overpic}
        \begin{overpic}[scale=0.20]{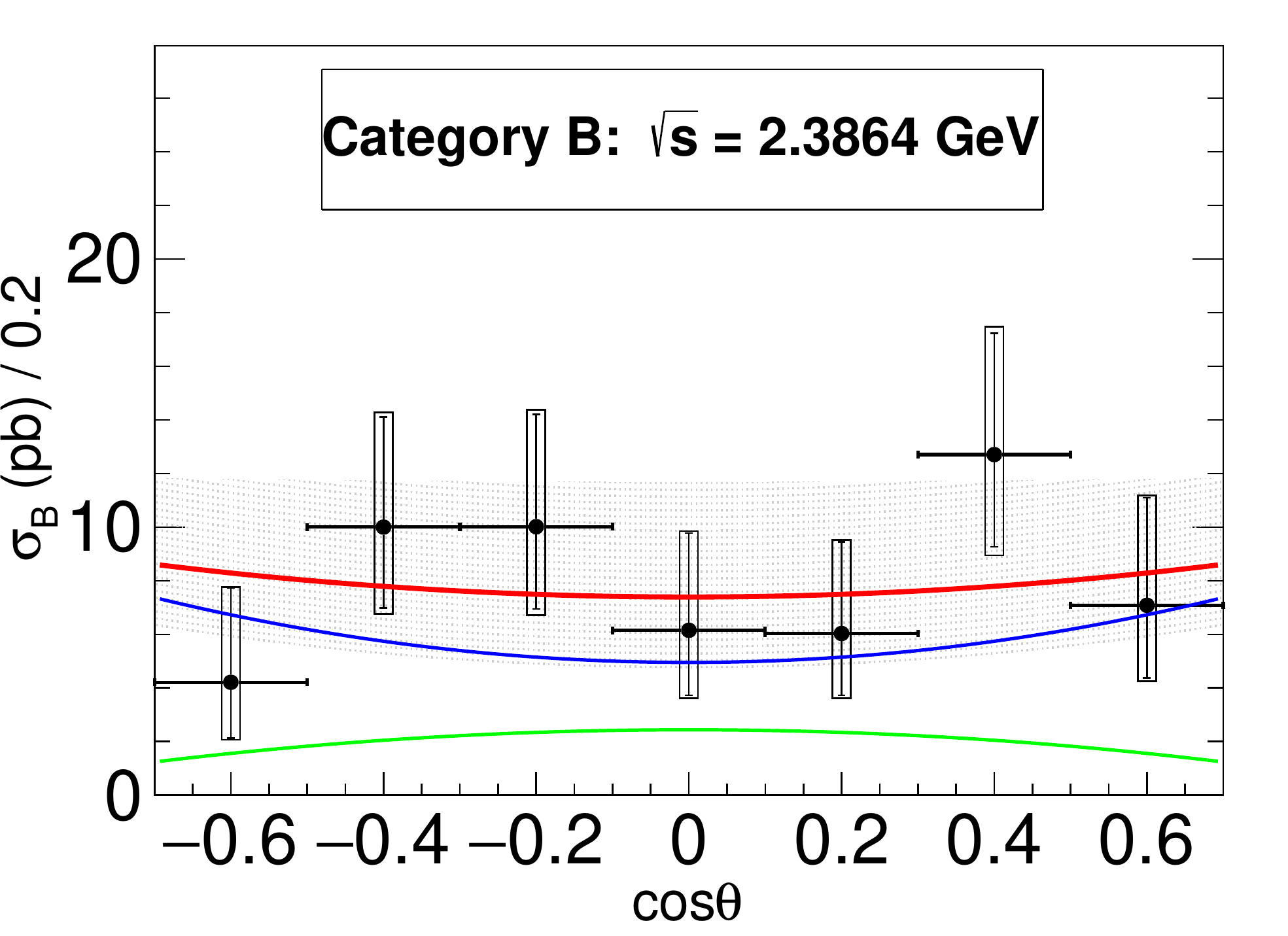}
        \end{overpic}
        \begin{overpic}[scale=0.20]{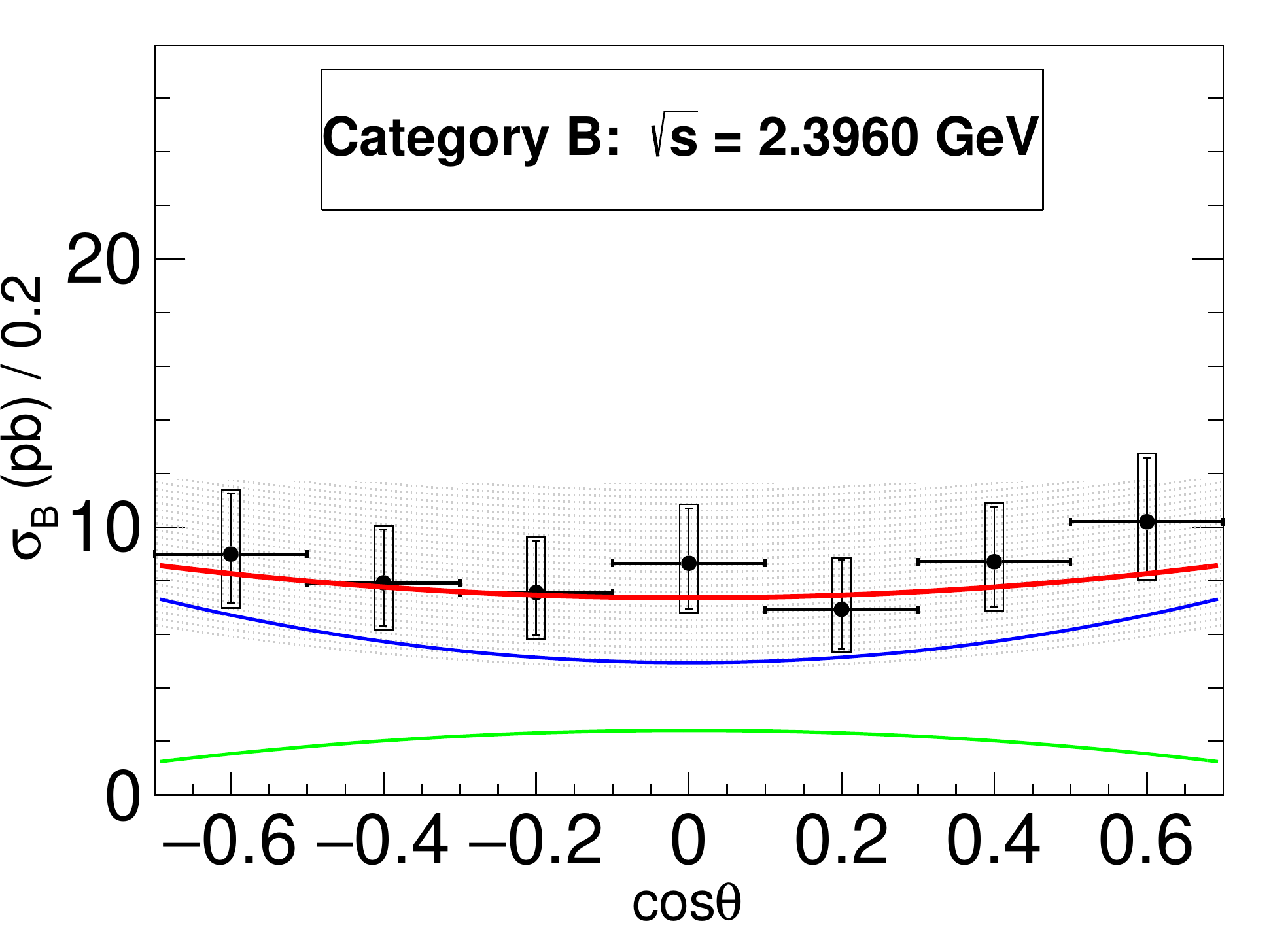}
        \end{overpic}
        \begin{overpic}[scale=0.20]{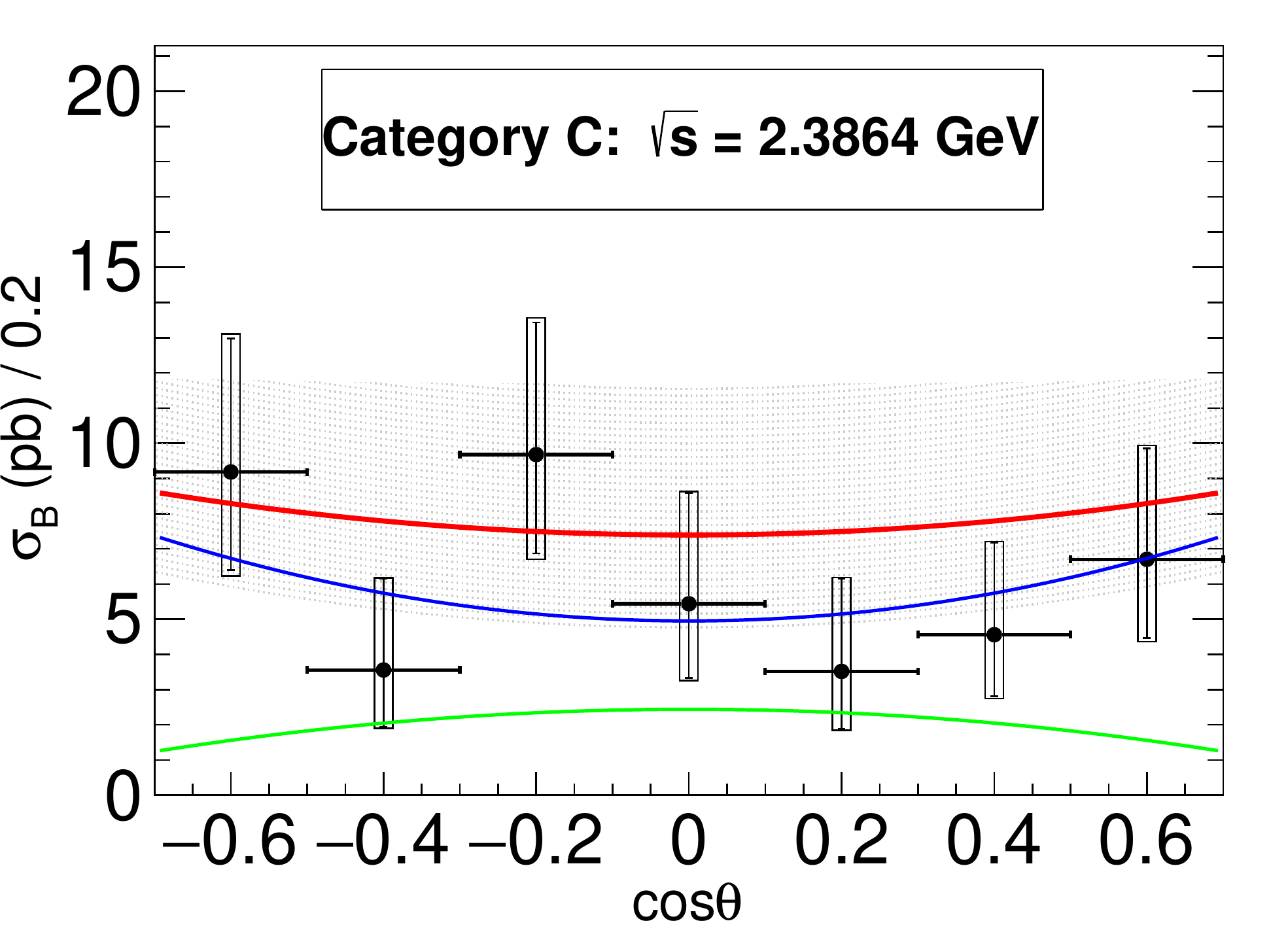}
        \end{overpic}
        \begin{overpic}[scale=0.20]{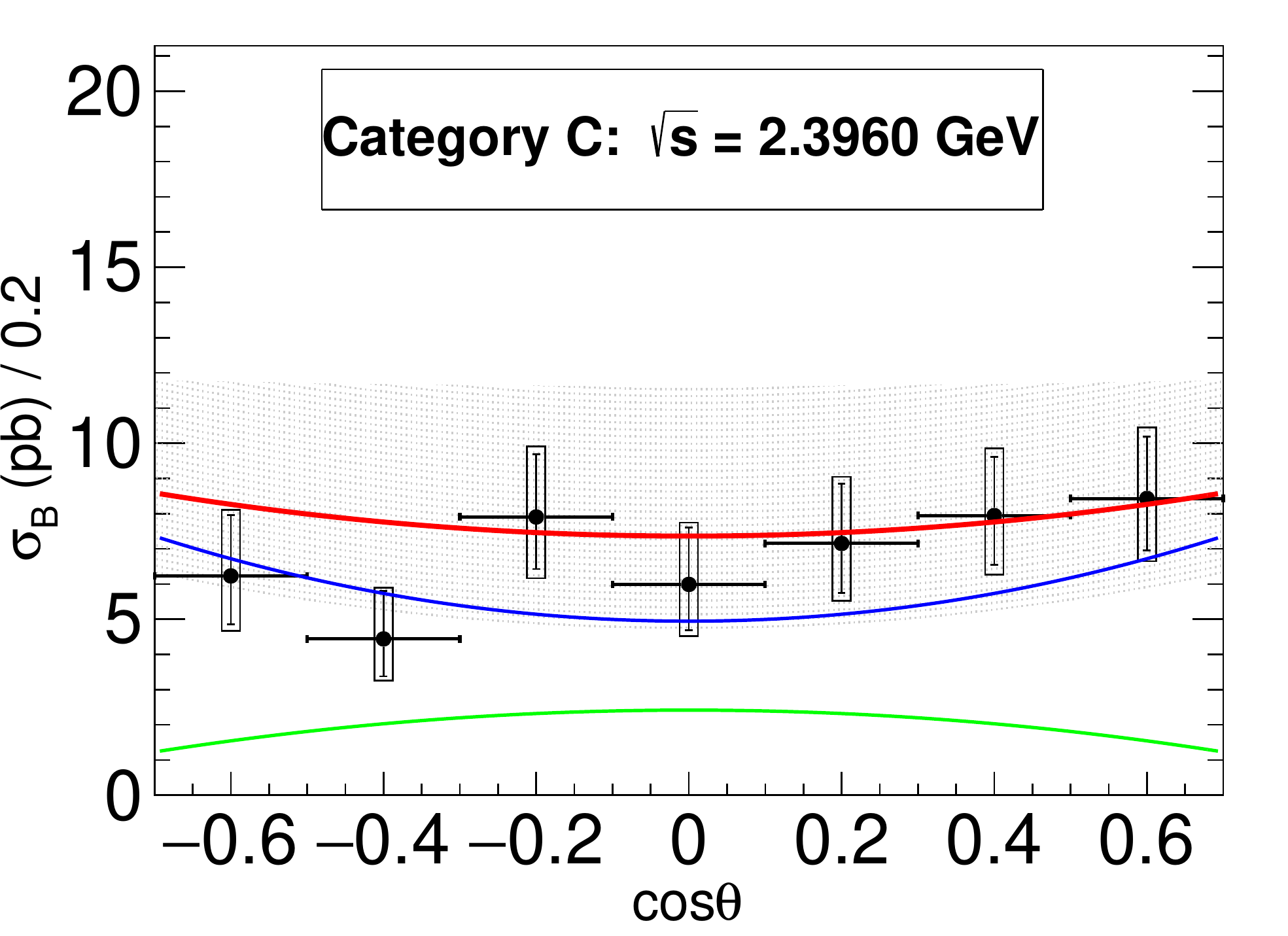}
        \end{overpic}\\
        %\vspace{0.2cm}
        \caption{A simultaneous fitting to the differential cross sections at $\sqrt{s}=2.3864$ and $2.3960$ GeV  for three event categories. Data are shown as black dots with error bars, the total fit as a red line and a gray band (68\% CI.), the $|G_M|^2$ component as the blue line, and the $\tau|G_E|^2$ component as the green line. The asymmetrical uncertainties for the data are determined from the Poisson distribution to take into account the low statistics in most bins. Total uncertainties are represented with boxes.
        }
        \label{1}
\end{figure}

With the above numbers for the three categories, $|G_M|$ and $R_{em}$ are determined by minimizing the negative logarithm of the likelihood function (NLL) based on the Poisson probability density function.
For example, Fig.~\ref{1} illustrates a simultaneous fitting to data of $\cos\theta_{\bar{n}}$-dependent cross sections at $\sqrt{s} = 2.3864$ and $2.396$ GeV.
The fitting results and associated parameters are summarized in Table~\ref{tab:angresults}.
Note that data collected at 12 c.m. energies is grouped into five c.m. energy intervals to maximize the statistical precision of the results.
The details on the fitting procedure and values of differential cross sections for the other energy points are listed in Ref.~\cite{supplementary}.

\begin{table}[h]
\scriptsize
   \tabcolsep=0.15cm
\renewcommand{\arraystretch}{1.4}
\begin{center}
\caption{ The integrated luminosity $\mathcal{L}$, form factor ratios $R_{em} = |G_E|/|G_M|$, electric $|G_E|$ and magnetic $|G_M|$ form factors. The first uncertainties are statistical and the second systematic.
The nominal energy for each energy interval is weighted by luminosity and c.m. energy of the corresponding data sample. The lower (upper) energy uncertainty is taken as the difference between the nominal energy and the lowest (highest) energy among the group.}
\begin{tabular}{c|cccc}
	\hline
	\hline
$\sqrt{s}$ (GeV) & $\mathcal{L}$ (pb$^{-1}$) &  $R_{em}$ & $|G_M|\ (\times 10^{-2})$ &  $|G_E|\ (\times 10^{-2})$  \\
	\hline
\makecell[c]{2.0000\\2.0500}               & \makecell[c]{10.1\\3.34}  & 0.9$\pm$0.7$\pm$0.4 &18.6$\pm$5.0$\pm$3.1 & 17.2$\pm$8.3$\pm$4.7  \\
\makecell[c]{2.1250\\2.1500}               & \makecell[c]{108\\2.84} & 1.3$\pm$0.4$\pm$0.3  & 8.7$\pm$1.2$\pm$0.8 & 11.2$\pm$1.7$\pm$1.1\\
\makecell[c]{2.1750\\2.2000\\2.2324\\2.3094} & \makecell[c]{10.6\\13.7\\11.9\\21.1}   &
1.5$\pm$0.6$\pm$0.2 & 6.5$\pm$1.5$\pm$0.4 &  9.8$\pm$1.9$\pm$0.6\\
\makecell[c]{2.3864\\2.3960}               & \makecell[c]{22.5\\66.9}   & 0.9$\pm$0.3$\pm$0.2 & 8.3$\pm$0.9$\pm$0.4 &  7.3$\pm$2.1$\pm$1.0\\
\makecell[c]{2.6454\\2.9500}               & \makecell[c]{67.7\\15.9}   & 0.6$\pm$0.9$\pm$0.7  & 4.4$\pm$0.8$\pm$0.3 &  2.5$\pm$2.9$\pm$2.9 \\
\hline
\hline
\end{tabular}
\label{tab:angresults}
\end{center}
\end{table}

Various sources of systematic uncertainties are considered for the determination of $|G_E|$, $|G_M|$, and $R_{em}$, including the uncertainties from the luminosity, the category-specific signal event selections, the MC model, the Born cross section input for the signal efficiency determination, the trigger efficiency, and the fit procedure. The uncertainty from the integrated luminosity is determined with large-angle Bhabha scattering to be $1\%$~\cite{bes3:lum1,bes3:lum2}. The uncertainty from the signal event selection is taken into account by using the efficiency corrections for the differences between data and signal MC. By varying $C^{dm}$ within $1\sigma$, the difference on the differential cross section is taken as the systematic uncertainty.
The uncertainty in the efficiency determination stemming from the form factor input model to the MC Born cross section  is reduced to 1\% by iterative efficiency determination and fitting. The uncertainty from the signal event extraction arises from three sources: the signal shape, the background shape, and the fitting range. By changing the signal and background shapes and varying the fitting range, the systematic uncertainty from the signal yield extraction is determined as the largest deviation from the nominal results. The uncertainty from the trigger efficiency is studied with a different parametrization of the detector response, as discussed in Ref.~\cite{np}. The overall systematic uncertainties at each bin are summarized in Ref.~\cite{supplementary}.
The variation of mean values of $R_{em}$ and $|G_M|$ with/without including systematic uncertainties during fitting angular distributions is taken as a systematic uncertainty.
In total, three categories are used to determine the final results of  $|R_{em}|$ and $|G_{M}|$ by taking into account correlations of systematic uncertainties at different bins.
The uncertainties of $|G_E|$ are propagated from uncertainties of $R_{em}$, $|G_M|$ and their correlations.
Table~\ref{tab:angresults} lists values of the total systematic uncertainties.

In conclusion,
values of $|G_E|$, $|G_M|$, and $R_{em}$
have been extracted at five c.m. energy intervals in the TL region.
The results for $R_{em}$
are consistent to unity
considering systematic uncertainties in a wide range of $q^2$.

\begin{figure}[h]
  \centering
  \includegraphics[width=0.36 \textwidth]{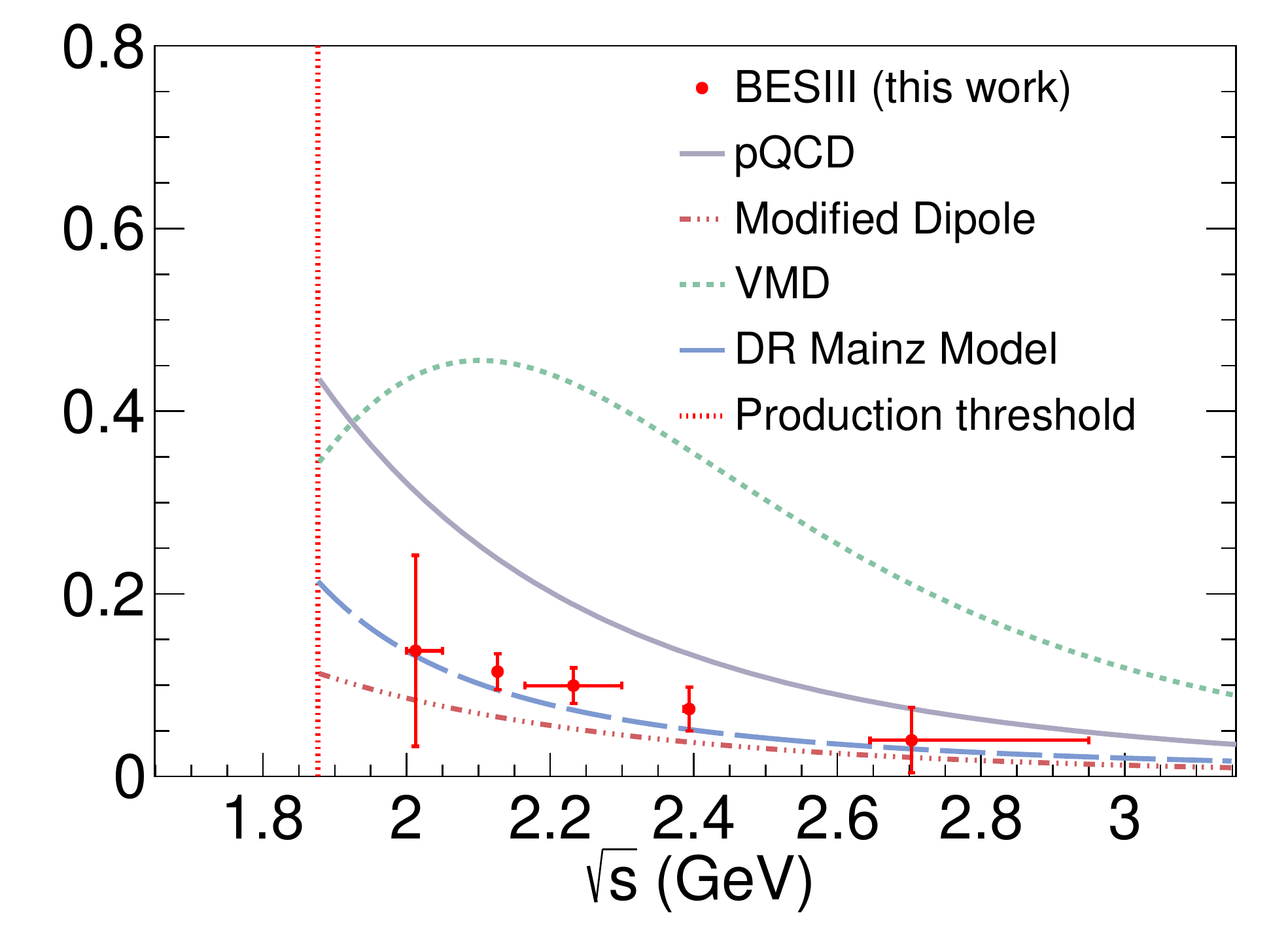}
  \includegraphics[width=0.36 \textwidth]{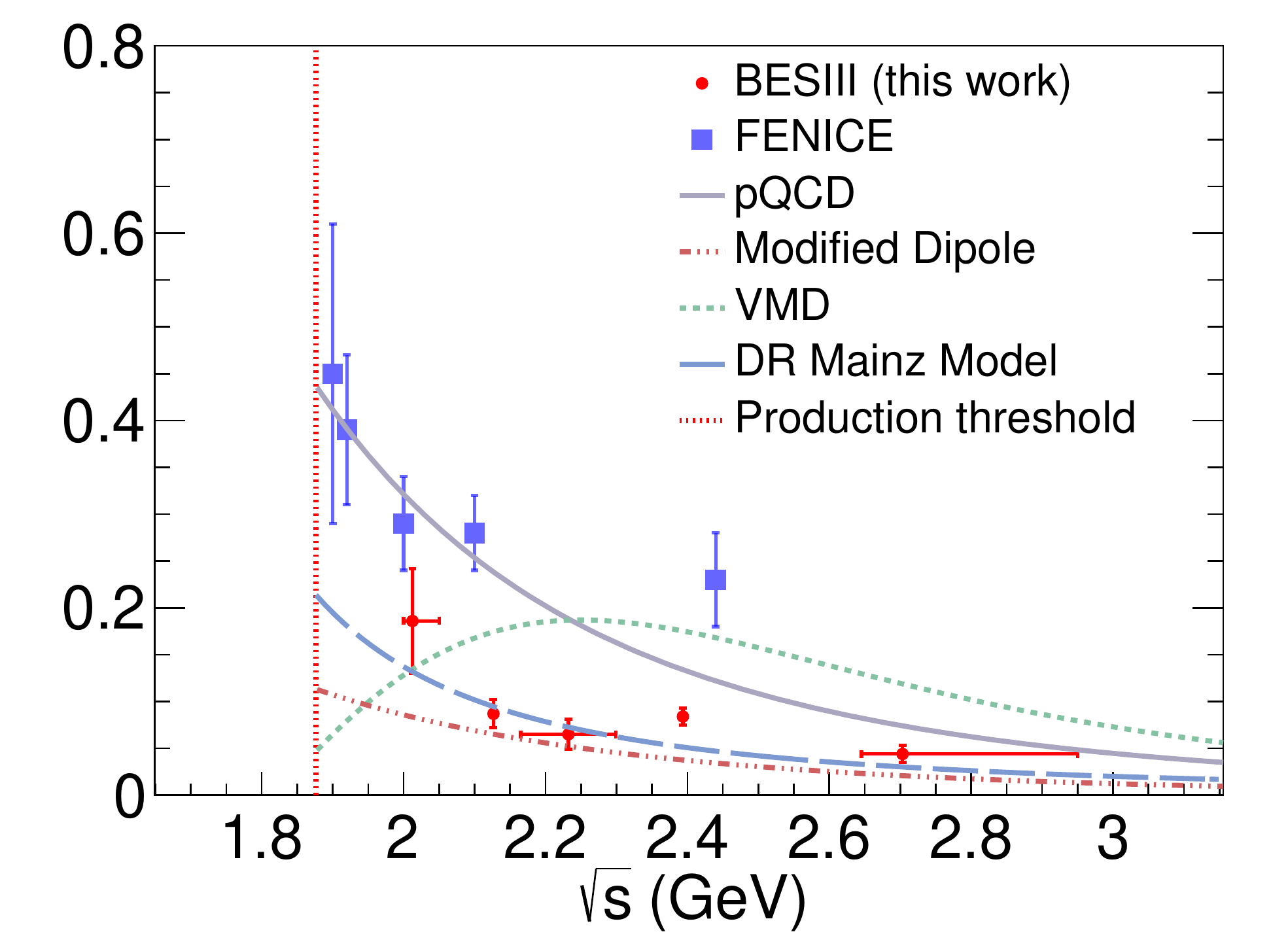}
   % \vspace{-0.25cm}
    \caption{
Results for the separated form factors of the neutron. (Top) Electric and (Bottom) magnetic form factors from this work (black dots with uncertainties) are shown together with results from the FENICE experiment~\cite{FENICE} extracted under the hypothesis $|G_E| = 0$ (blue rectangles) and four different parametrizations~\cite{Shirkov:1997wi,Bianconi:2015owa,Hammer:1996kx}.
 The vertical red dotted line indicates the production threshold.}
  \label{fig:pqcdvmdpara}
\end{figure}

Compared with the FENICE results, the values for $|G_M|$ from this work are smaller by a factor of $\sim$ 2-3 in the range of $\sqrt{s} = 2.0\sim2.5$ GeV, as shown in Fig.~\ref{fig:pqcdvmdpara}.
The measured $|G_E|$ and $|G_M|$ can be used to test various nucleon models
to provide a more comprehensive picture of the nucleon structure.
Among models such as a parametrization obtained from the pQCD~\cite{Shirkov:1997wi}, a modified dipole model based on the quark counting rule and analytical extension (MD)~\cite{Tomasi-Gustafsson:2001wyw},
a Vector Meson Dominance model (VMD)~\cite{VMD}, and a model based on Dispersion Relations (DR) \cite{Hammer:1996kx,Lin:2021umz,Lin:2021xrc}, our results show the best agreement with the DR-based model (long-dashed line).
Note that the modified dipole parametrization (dot-dashed line)  is re-analyzed with the experimental results from this work.
The free parameters of the DR-based model are optimized by a fit to the TL $|G_M|$ data, which are extracted for the neutron under the hypothesis that $|G_E|=|G_M|$. The free parameters of the MD model are optimized with a fit to the TL effective FFs. The pQCD based parametrization was initially developed for $|G_E|$ and $|G_M|$, legitimizing the use of these models also for a comparison with $|G_E|$.
In contrast, the VMD model predicts different values for $|G_E|$ and $|G_M|$.

\begin{figure}[h]
  \centering
\includegraphics[width=0.36 \textwidth]{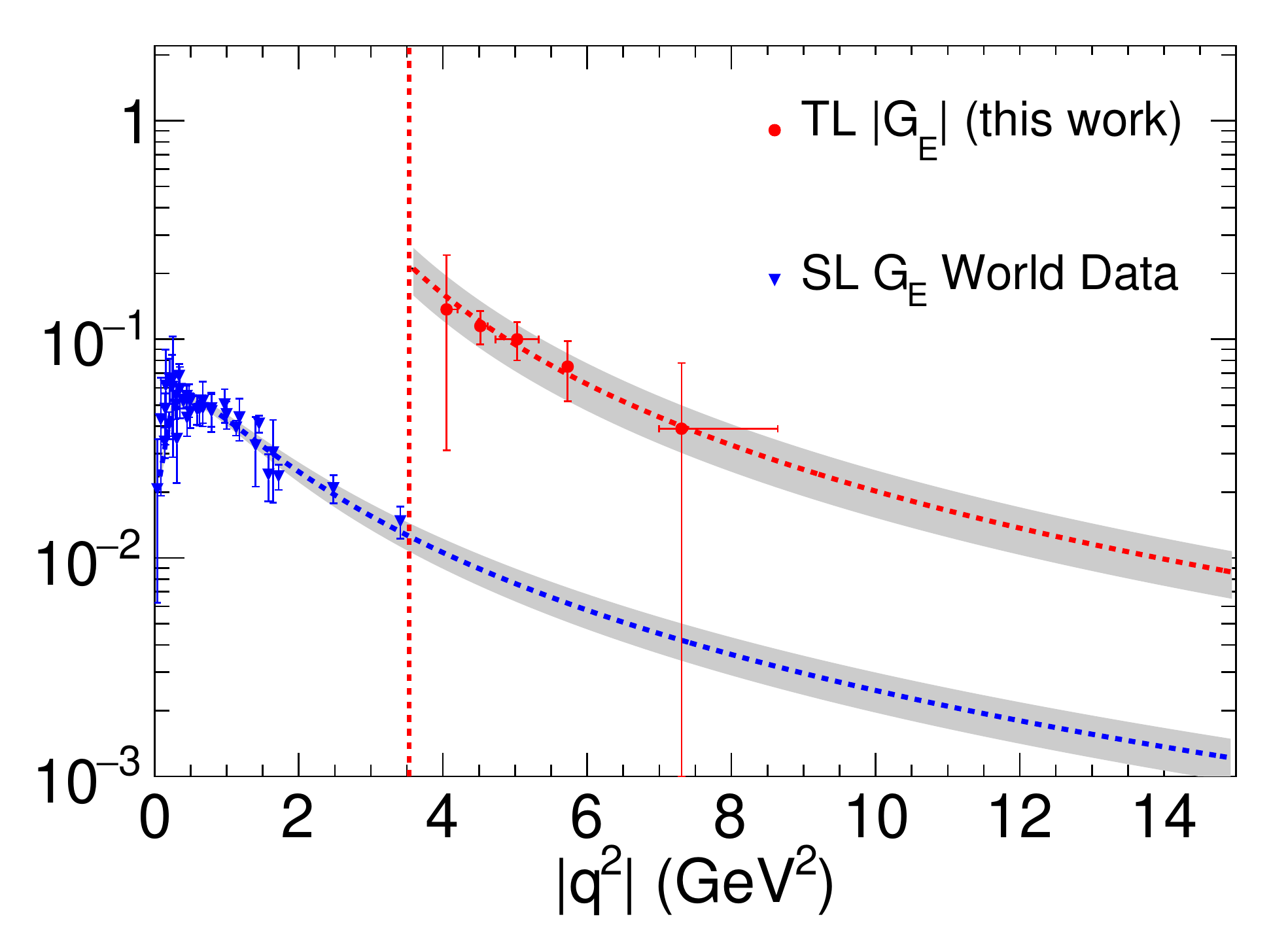}
\includegraphics[width=0.36 \textwidth]{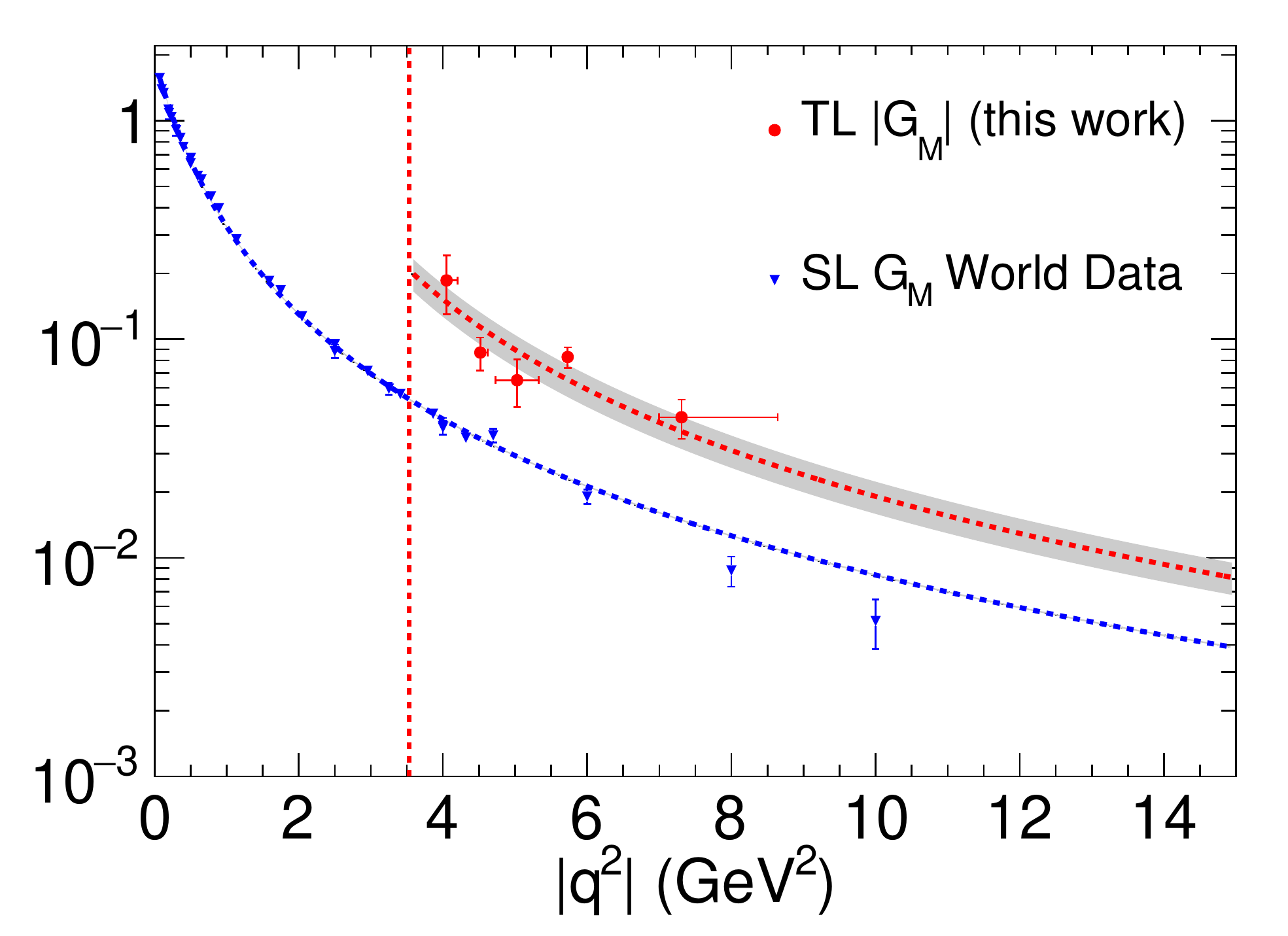}
   % \vspace{-0.25cm}
    \caption{
Testing analyticity of
(Top) Electric and (Bottom) Magnetic form factors from this work (red dots with uncertainties) shown together with results from the world data of SL ones.
The fit to TL (SL) data is represented with a red (blue) line and a gray band (95\% CI.).
The related formulas are listed in table~\ref{tab:tlsl_fitting_results2}.
The vertical red dotted line indicates the production threshold.}
  \label{fig:gegm1}
\end{figure}

\begin{figure}[h]
  \centering
    \includegraphics[width=0.36 \textwidth]{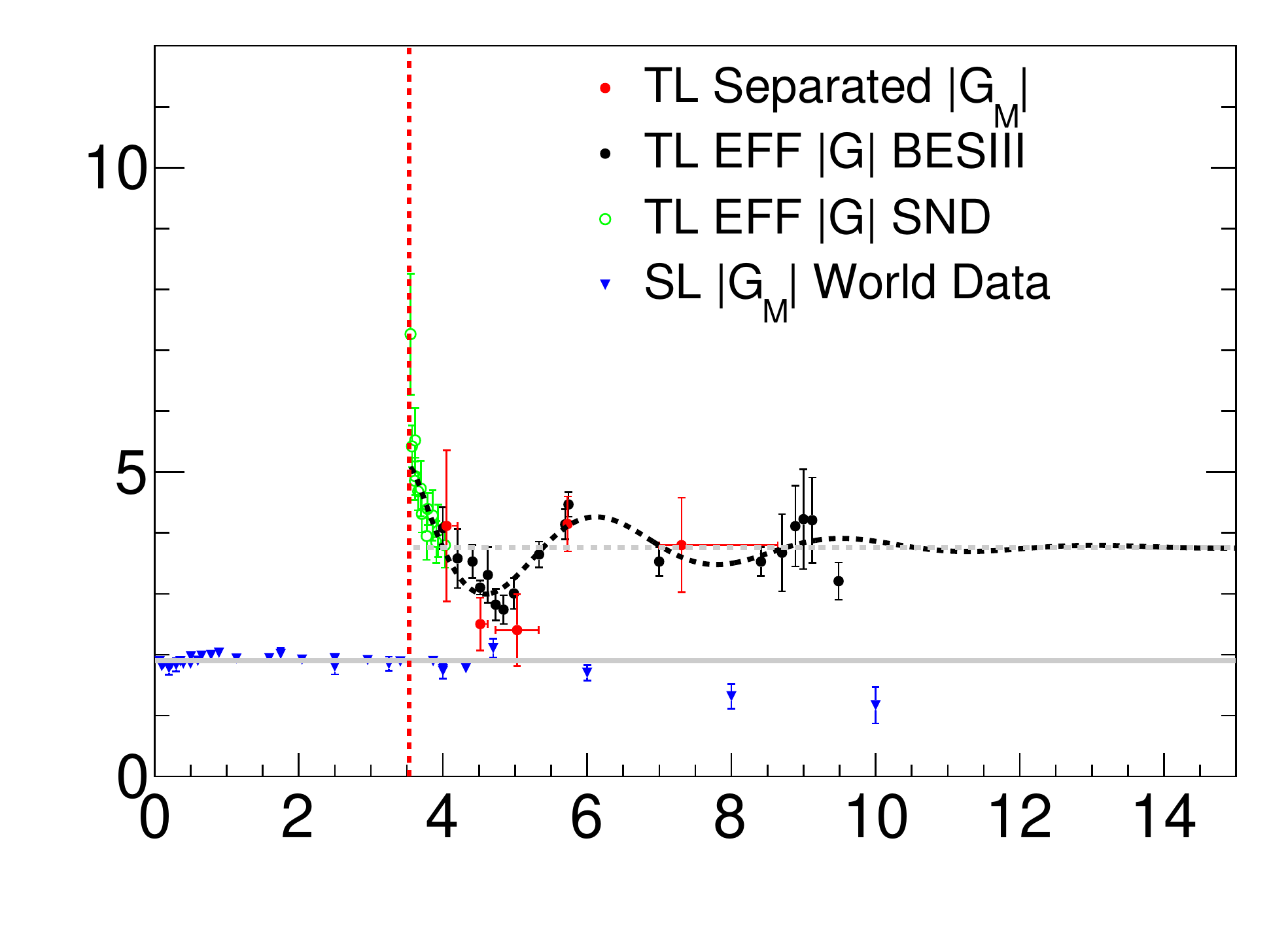}
    \includegraphics[width=0.36 \textwidth]{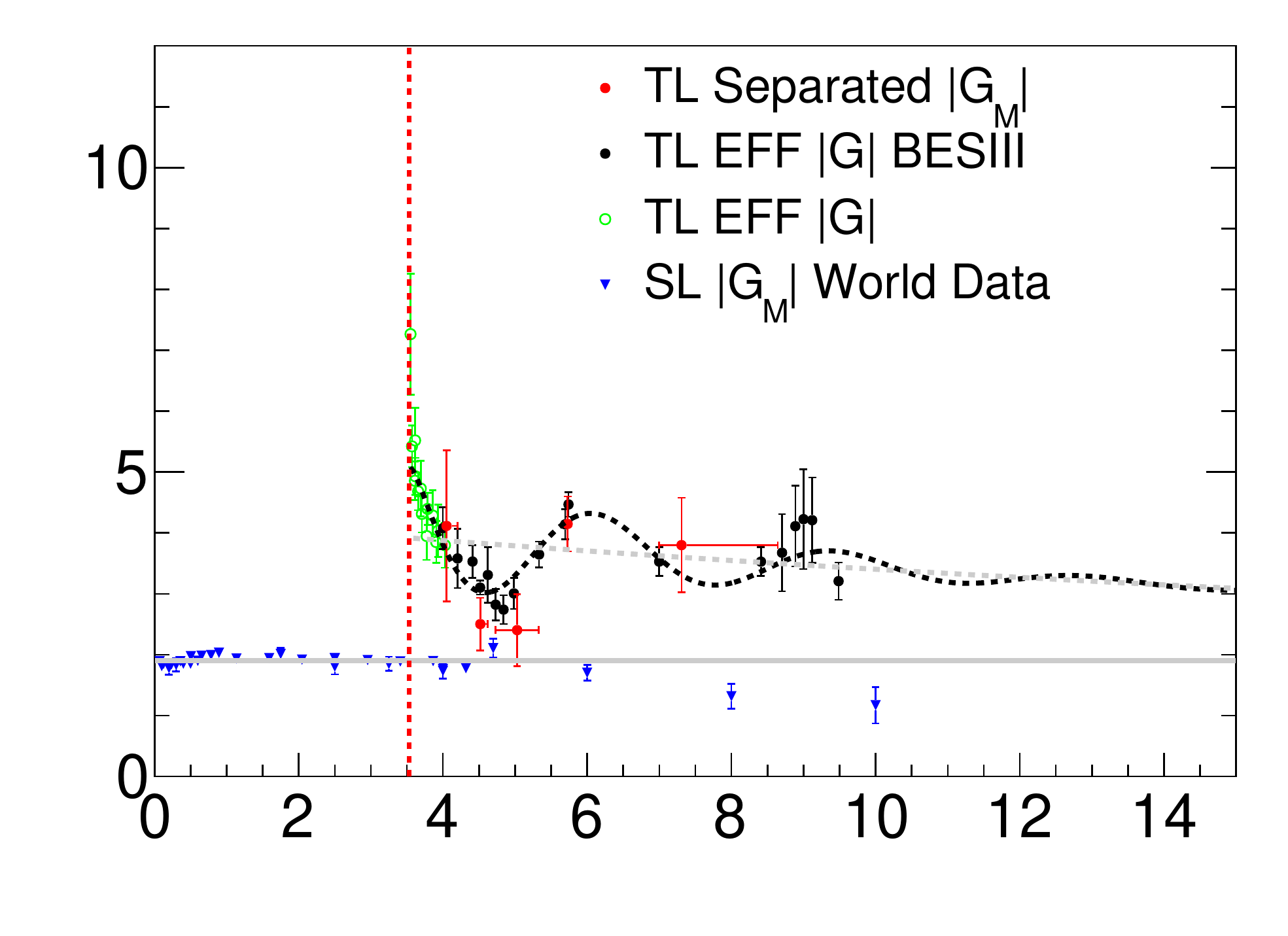}
    \includegraphics[width=0.36 \textwidth]{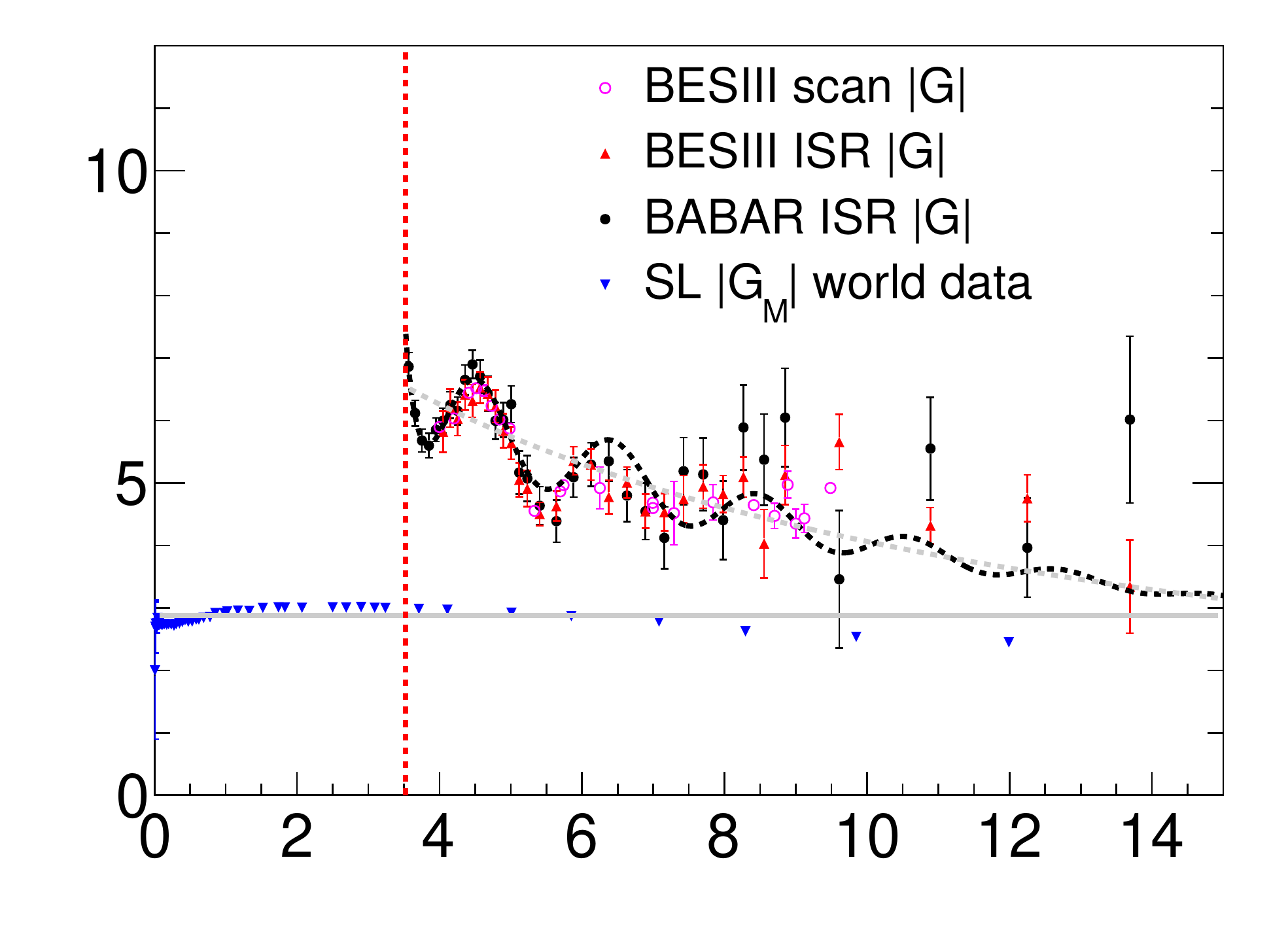}
   % \vspace{-0.25cm}
    \caption{The world data set of magnetic form factors in the SL region (blue down triangles).
    The full gray lines represent the formula $\mathcal{F}_0$.
 (Top) measurements of the neutron effective factors from  BESIII (black) and SND (green) in the TL region.
 The dashed black lines represent the formula $\mathcal{F}_1$ in Ref.~\cite{supplementary}, and the dashed gray lines represent the formula $\mathcal{F}_0$.
 (Middle) The dashed black lines represent the formula $\mathcal{F}_3$,
 and the dashed gray lines represent the formula $\mathcal{F}_2$.
 (Bottom) measurements of the proton effective form factors from BESIII (hollow circle, triangle) and BABAR (solid circle) in the TL region. The dashed black lines represent the formula $\mathcal{F}_3$,
 and the dashed gray lines represent the formula $\mathcal{F}_2$.
}
  \label{fig:gegm2}
\end{figure}

The EMFFs derived from data of unpolarized experiments empirically scale like $G_{E,M} \sim (-q^2)^{-2}$~\cite{pQCDcount2} in case of $q^2\to -\infty$ in the SL region.
It is interesting to check whether the TL form factors show any asymptotic behavior.
Our results show $|G_{E,M}^n|\sim (q^2)^{-2}$ in the TL region.
It is important to test the analyticity of EMFFs as a direct
consequence of micro causality and unitarity.
As stated in the Phragm\`{e}n-Lindel\"{o}f (P-L) theorem~\cite{PLtheo}, EMFFs in the TL region can be extended to any direction of the $q^2$ complex plane. As a result, the numerical values of EMFFs should approach each other for $|q^2| \rightarrow \infty$,
i.e., $\mathcal{R}^{E,M}\equiv|\frac{G_{E,M}^{TL}(q^2)}{G_{E,M}^{SL}(-q^2)}|\xrightarrow{|q^2|\to\infty} 1$.
Fig.~\ref{fig:gegm1} shows that
the TL $|G_E|$ ($|G_M|$) has no intersections with the SL $G_E$ ($G_M$),
using an extrapolation with current fitting parameters
for the neutron. The measured ratios are $\mathcal{R}^{E}=5.18\pm1.18 > 1$ for the electric form factors
and $\mathcal{R}^{M}=1.72\pm0.14 > 1$ for the magnetic form factors.
We further check the analyticity of effective form factors  $|G|$
as shown in Fig.~\ref{fig:gegm2}.
The neutron TL $|G_{n}|$ have no intersection with its SL $G_M^n$ multiplied by its magnetic moment $\mu_{n}$ when following the parameterized formula $\mathcal{F}_1$ in Ref.~\cite{supplementary}.
Note that the illustrated fitting (black dot-dashed line) makes use of data of effective FFs from both BESIII (black dots) and SND (green circles).
A refit with the parameterized formula $\mathcal{F}_3$
%in Ref.~\cite{supplementary}
shows an intersection around 47$\pm$41 GeV$^2$.
The proton TL $|G_{p}|$,
which have been normalized to the reciprocal of a dipole FF, has an intersection around 18$\pm$6 GeV$^2$ with
its SL $G_M^p$ multiplied by its magnetic moment $\mu_{p}$ when using the parameterized formula $\mathcal{F}_3$.
%in Ref.~\cite{supplementary}.
Note that the illustrated fitting (black dot-dashed line) used only BABAR data.
The related fitting results are listed in table
~\ref{tab:tlsl_fitting_results2} and ~\ref{tab:tlsl_fitting_results}.
%in Ref.~\cite{supplementary}.

\begin{table}[!h]
\scriptsize
   \tabcolsep=0.15cm
\renewcommand{\arraystretch}{1.4}
\begin{center}
\caption{Fitting results corresponding to Fig.~\ref{fig:gegm1}.
Numbers in top 5 rows corresponds to the top plot,
numbers in bottom 5 rows corresponds to the bottom plot.
}
\begin{tabular}{c|cc}
	\hline
	\hline
			\multicolumn{3}{c}{$|G_E|$} \\
	\hline
     -     & TL ($q^2>0$) &  SL ($q^2<0$) \\
	\hline
   formula         &  $\frac{A}{(1-q^2/0.71)^2}$  & $\frac{A\tau}{(1+B\tau)}\frac{1}{(1-q^2/0.71)^2}, \tau=-q^2/4m_n^2$     \\
   parameters      & A=3.39$\pm$0.43  &  A=1.42$\pm$0.08, B=$2.17\pm0.39$               \\
   $\chi^2$/ndf   & 0.4/4          &  25/36   \\
	\hline
	\multicolumn{3}{c}{$|G_M|$} \\
	\hline
     -     & TL ($q^2>0$) &  SL ($q^2<0$)\\
	\hline
   formula         &  $\frac{A}{(1-q^2/0.71)^2}$  & $\frac{A}{(1-q^2/0.71)^2}$     \\
   parameters      & A=3.27$\pm$0.28  &  A=1.899$\pm$0.008               \\
   $\chi^2$/ndf   & 8.8/4          &  82/31   \\
	\hline

\hline
\end{tabular}
\label{tab:tlsl_fitting_results2}
\end{center}
\end{table}

\begin{table}[t]
\scriptsize
   \tabcolsep=0.15cm
\renewcommand{\arraystretch}{1.4}
\begin{center}
\caption{Fitting results corresponding to Fig.~\ref{fig:gegm2},
the second column to bottom, the third column to middle, and the fourth column to top.}
\begin{tabular}{c|ccc}
	\hline
	\hline
     parameters     & proton ($  \mathcal{F}_1$)  & neutron ($  \mathcal{F}_1$) & neutron ($  \mathcal{F}_2$) \\
	\hline
   A         & 9.79$\pm$0.78  & 4.27$\pm$0.32  &3.76$\pm$0.07   \\
 $m_a^2$     & 7.08$\pm$1.43  &39.19$\pm$25.40 & -              \\
$A_{osc}$    & 0.06$\pm$0.01  & 0.08$\pm$0.02  &0.10$\pm$0.02   \\
   B         & 0.88$\pm$0.19  & 0.98$\pm$0.18  &1.15$\pm$0.18   \\
   C         & 5.40$\pm$0.17  & 3.32$\pm$0.21  &3.27$\pm$0.23   \\
   D         & 6.33$\pm$0.19  & 5.50$\pm$0.29  &5.47$\pm$0.28   \\
$\chi^2$/ndf &   38.9/31      &  27.5/27       &30.9/27         \\
	\hline
\hline
\end{tabular}
\label{tab:tlsl_fitting_results}
\end{center}
\end{table}

In summary, we have separated $|G_E|$ from $|G_M|$ for the neutron within a wide range of $q^2$ from 4 to  9 GeV$^2$ with relative uncertainty around 12\% for the modulus of the magnetic form factor.
This is comparable in accuracy to results from electron scattering in a similar SL region of four-momentum transfer.
In the future, further efforts will be made not only at electron accelerators~\cite{Accardi:2012qut,eicc}
but also at electron-positron~\cite{BESIII:2020nme} and proton-antiproton colliders~\cite{panda} to obtain a global picture of all data in the TL and SL regions which will further deepen our understanding of the nucleon structure.

\vspace{0.25cm}
%% Saved at => 2022-10-11
%\textbf{Acknowledgement}
The BESIII collaboration thanks the staff of BEPCII and the IHEP computing center for their strong support. This work is supported in part by National Key R\&D Program of China under Contracts Nos.~2020YFA0406400, 2020YFA0406300; National Natural Science Foundation of China (NSFC) under Contracts Nos.~11635010, 11735014, 11805124, 11835012, 11935015, 11935016, 11935018, 11961141012, 12022510, 12025502, 12035009, 12035013, 12061131003, 12122509, 12192260, 12192261, 12192262, 12192263, 12192264, 12192265, 12221005; the Chinese Academy of Sciences (CAS) Large-Scale Scientific Facility Program; the CAS Center for Excellence in Particle Physics (CCEPP); Joint Large-Scale Scientific Facility Funds of the NSFC and CAS under Contract No.~U1832207; CAS Key Research Program of Frontier Sciences under Contracts Nos.~QYZDJ-SSW-SLH003, QYZDJ-SSW-SLH040; 100 Talents Program of CAS;
Guangdong Major Project of Basic and Applied Basic Research no.~2020B0301030008; Science and Technology Program of Guangzhou (no.~2019050001);
The Double First-Class university project foundation of USTC;
The Institute of Nuclear and Particle Physics (INPAC) and Shanghai Key Laboratory for Particle Physics and Cosmology; ERC under Contract No.~758462; European Union's Horizon 2020 research and innovation programme under Marie Sklodowska-Curie grant agreement under Contract No.~894790; German Research Foundation DFG under Contracts Nos.~443159800, 455635585, Collaborative Research Center CRC 1044, FOR5327, GRK 2149; Istituto Nazionale di Fisica Nucleare, Italy; Ministry of Development of Turkey under Contract No.~DPT2006K-120470; National Research Foundation of Korea under Contract No. NRF-2022R1A2C1092335; National Science and Technology fund; National Science Research and Innovation Fund (NSRF) via the Program Management Unit for Human Resources \& Institutional Development, Research and Innovation under Contract No.~B16F640076; Polish National Science Centre under Contract No.~2019/35/O/ST2/02907; Suranaree University of Technology (SUT), Thailand Science Research and Innovation (TSRI), and National Science Research and Innovation Fund (NSRF) under Contract No.~160355; The Royal Society, UK under Contract No.~DH160214; The Swedish Research Council; U. S. Department of Energy under Contract No.~DE-FG02-05ER41374.

%\textbf{Other Fund Information}
%To be inserted with an additional sentence into papers that are relevant to the topic of special funding for specific topics. Authors can suggest which to Li Weiguo and/or the physics coordinator.
%Example added sentence: This paper is also supported by the NSFC under Contract Nos. 10805053, 10979059, ....National Natural Science Foundation of China (NSFC), 10805053, PWANational Natural Science Foundation of China (NSFC), 10979059, Lund弦碎裂强子化模型及其通用强子产生器研究National Natural Science Foundation of China (NSFC), 10775075, National Natural Science Foundation of China (NSFC), 10979012, baryonsNational Natural Science Foundation of China (NSFC), 10979038, charmoniumNational Natural Science Foundation of China (NSFC), 10905034, psi(2S)->B BbarNational Natural Science Foundation of China (NSFC), 10975093, D 介子National Natural Science Foundation of China (NSFC), 10979033, psi(2S)->VPNational Natural Science Foundation of China (NSFC), 10979058, hcNational Natural Science Foundation of China (NSFC), 10975143, charmonium rare decays
%% ends here %%

%%%%%%%%%%%%%%%%%%%%%%%%%%%

\clearpage
\onecolumngrid
\textbf{\Large Supplemental Material for ``Measurements of the electric and magnetic form factors of the neutron for time-like momentum transfer}
\appendix
\setcounter{table}{1}
\setcounter{figure}{5}

\begin{appendices}

\section{MAXIMUM LIKELIHOOD FITTING}
\label{sec_supp_sysB}
\hypertarget{item:d}{}

The predicted cross-section $\sigma^{pred}_i$ at {\it{i-th}} bin for each category X (=A,B,C)  can be obtained as below:
 \begin{eqnarray}\label{eqnxs_formula4}
\sigma_{i}^{pred} & =  & \int_{bin}  \frac{d\sigma}{d\cos\theta} d\cos\theta,
 \end{eqnarray}
with the theoretical formula,
 \begin{equation}
 \frac{d\sigma}{d\cos\theta} =  \frac{\pi\alpha^2\beta}{2s} |G_M|^2(1+\cos^2\theta + \tau R_{em}^2\sin^2\theta),
\label{eqnxs_formula3}
\end{equation}
where  $\tau = \frac{4m_n^2}{s}$, $|G_M|$ and $R_{em}\equiv |G_E|/|G_M|$ are two free parameters shared by 3 categories to be determined with the following minimization.

We minimize a negative logarithm likelihood as below:
\begin{eqnarray}
\label{eqnxs_formula5}
     LL & =& -log(\prod_{X=A,B,C}\prod_{i=1}^{7} \frac{(N^{pred}_{X,i})^{N^{obs}_{X,i}} e^{-N^{pred}_{X,i}} }{N^{obs}_{X,i} !}) \\
     N^{pred}_{X,i} & = &\left\{
     \begin{array}{l}
\tiny      \mathcal{L}\sigma^{pred}(\mathcal{E}^{MC} C_{dm}C_{trg}(1+\delta))_{A,i},  \\
\tiny       \mathcal{L}\sigma^{pred}(\mathcal{E}^{MC} C_{dm}C_{trg}C_{muc}(1+\delta))_{B,i}, \\
\tiny       \mathcal{L}\sigma^{pred}(\mathcal{E}^{MC} C_{dm}C_{trg}C_{ee}C_{muc}(1+\delta))_{C,i}, \\
      \end{array}
      \right.
\end{eqnarray}
where $N^{obs}_{X,i}$ and $N^{pred}_{X,i}$ represent number of observed events and number of predicted events, $\mathcal{L}$ is the luminosity,
$\mathcal{E}^{MC}$ is the signal efficiency, ($C_{dm}$, $C_{trg}$, $C_{ee}$, $C_{muc}$) are four types of efficiency corrections,
($1+\delta$) is the initial-state-radiation and vacuum polarization correction determined by the signal MC generator, at $i$-th bin for category X.

The systematic uncertainty at each bin is calculated to be
$\Delta_{T,i} = \sqrt{ \Delta^2_{uc,i} + \Delta^2_{c,i}}$,
where $\Delta_{uc}= N^{obs}*\sqrt{R_{dm}^2 + R_{isr}^2}$ represents the uncorrelated uncertainty,
and  $\Delta_{c} = N^{obs}*R_{c}$ the correlated one. Here, $R_c$ is the relative systematic uncertainties which are
quoted from Ref.~\cite{np}. $R_{dm}$ is directly calculated with $C_{dm}$ values, for example $C_{dm}=1.23\pm0.10$ yields a $R_{dm}=\frac{0.10}{1.23}$,
so are the others.

Thus the negative logarithm likelihood becomes:
\begin{eqnarray}
\label{eqnxs_formula5}\scriptsize
     LL & =& -log(\prod_{X=A,B,C}\prod_{i=1}^{7} \int P_a \otimes P_b), \\
    P_a & = & \frac{(N^{pred}_{X,i})^{N^{obs}_{X,i}} e^{-N^{pred}_{X,i}} }{N^{obs}_{X,i}\ !}, \\
     P_b & = & e^{-\frac{1}{2} (\frac{N^{obs}_{X,i}-N^{pred}_{X,i}}{N^{obs}_{X,i}*R_{X,i}})^2 },
\end{eqnarray}
using a convolution between a Poisson probability ($P_a$) and a Gaussian probability ($P_b$) which describes the systematic uncertainties per bin per category.

\section{FITTING ELECTROMAGNETIC FORM FACTORS}
\hypertarget{item:e}{}
\label{sec_supp_sysE}

Figures 6 illustrates the fitting to data of the effective form factors with the following formulas~\cite{Bianconi:2015owa,np}:

\begin{equation}
\begin{split}\label{eqn:formula1}
 \mathcal{F}_0  =    A, \quad
 \mathcal{F}_1  =    A
         +  A_{osc} \exp(-B \gamma\beta\sqrt{q^2})
            \times \cos(C\gamma\beta\sqrt{q^2} +D) (1-q^2/0.71)^2,\\
%\end{split}
%\end{equation}
%\begin{equation}
%\label{eqn:formula2}
%\begin{split}
\mathcal{F}_2 =    \frac{A}{1+q^2/m_a^2}, \quad
  \mathcal{F}_3  =    \frac{A}{1+q^2/m_a^2}
         +  A_{osc} \exp(-B \gamma\beta\sqrt{q^2})
            \times \cos(C\gamma\beta\sqrt{q^2} +D) (1-q^2/0.71)^2.
\end{split}
\end{equation}

\section{SYSTEMATIC UNCERTAINTIES}
\hypertarget{item:g}{}
Tables~\ref{tab:syserrors}, ~\ref{tab2:sumsys} and~\ref{tab3:sysall} lists systematic uncertainties of antineutron angular distributions per $\cos\theta_{\bar{n}}$ bin
at 12 c.m. energies for three categories:
luminosity $(U^A_L)$, antineutron and neutron selections $(U^A_{n,\bar{n}})$, $T_{0}$ and $T_0^n$ misalignment $(U^A_{T_0})$, angular distribution $(U^A_{model})$, antineutron and neutron misidentification ($U^A_{mis}$), time resolution of the neutron $(U^A_{res})$, event-level selections $(U^A_{evt})$, signal yields extraction $(U^A_{fit})$, Conexc generator iteration $(U^{A}_{ISR})$, trigger efficiency $(U^A_{tri})$.
Table~\ref{tab:angresults3} lists values of
$R_{em} = |G_E|/|G_M|$, electric $|G_E|$ and magnetic $|G_M|$ form factors together with their systematic uncertainties.

Two steps are performed to extract the free parameters, and their uncertainties. Firstly angular distributions of the antineutron are analyzed only considering the statistical uncertainties per bin. The values of $R_{em}$ and $|$$G_M$$|$ can be estimated directly using the open source packages: RooFit and TMinuit. Secondly angular distributions of the antineutron including the systematic uncertainties per bin are further analyzed. The values of $R_{em}$ and $G_M$ are obtained and compared to previous ones. The differences between central values obtained with/without including systematic uncertainties of angular distributions of the antineutron is also considered as a source of systematic uncertainty as summarized in Table~\ref{tab:angresults3}.
From these numbers, the statistical uncertainties are dominant and the systematic uncertainties are much smaller than the statistical uncertainties, because
main sources of systematic uncertainties are correlated among bins.
Considering the relation $\sigma\propto$ $|$G$_M$$|$$^2$ $ [({1+ \tau R_{em}^2}) +(1-\tau R_{em}^2) \cos^2\theta_{\bar{n}}]$, the power 2 means uncertainties of form factors are half discounted compared to uncertainties of cross sections of the process $e^+e^-\to n\bar{n}$ after the error propagation.
Since $|$$G_E$$|$ is strongly anti-correlated to $|$$G_M$$|$, the uncertainties of  $|$$G_E$$|$ are estimated with both uncertainties of  $|$$G_M$$|$ and uncertainties of $R_{em}$, and values of the correlation coefficient between $|$$G_M$$|$ and $R_{em}$ in Table~\ref{tab:angresults3} in the error propagation.
(Ref: https://root.cern/download/doc/RooFit\_Users\_Manual\_2.91-33.pdf)

\begin{table*}[!h]
\begin{center}
\footnotesize
\caption{Summary of all systematic uncertainty sources in Category A. The last column is the summary of the total uncertainties.}
\begin{tabular}{l|cccccccccccc}
  \hline
  \hline
  $\sqrt{s}~\gev$ & $U^A_L$  & $U^A_{n,\bar{n}}$ & $U^A_{T_0}$ & $U^A_{model}$ &  $U^A_{mis}$ &$U^A_{res}$ & $U^A_{fit}$  & $U^{A}_{ISR1}$ & $U^{A}_{ISR2}$ &  $U^A_{tri}$ & Total (\%)  \\
\hline
2.0000   &  1.0 & 3.1 &   0.2  & 10.8  &  0.1   &   5.0  &  4.2   & 1.0  & 0.3  & 4.6& 13.9      \\
2.0500   &  1.0 & 2.7 &   0.2  & 10.3  &  0.1   &   5.0  &  3.5   & 1.0  & 0.6  & 4.6& 13.2      \\
2.1000   &  1.0 & 2.4 &   0.2  &  7.8  &  0.1   &   5.0  & 12.1   & 1.0  & 1.2  & 4.6& 16.2      \\
2.1500   &  1.0 & 2.3 &   0.2  &  9.5  &  0.1   &   5.0  &  4.0   & 1.0  & 1.4  & 4.6& 12.7      \\
2.1250   &  1.0 & 2.3 &   0.2  &  8.1  &  0.1   &   5.0  & 12.3   & 1.0  & 1.7  & 4.6& 16.5       \\
2.1750   &  1.0 & 2.2 &   0.2  &  8.8  &  0.1   &   5.0  & 10.9   & 1.0  & 1.8  & 4.6& 15.9      \\
2.2000   &  1.0 & 2.1 &   0.2  &  8.0  &  0.1   &   5.0  &  8.3   & 1.0  & 2.2  & 4.5& 13.8       \\
2.2324   &  1.0 & 2.1 &   0.2  &  8.0  &  0.1   &   5.0  &  8.8   & 1.0  & 2.2  & 4.5& 14.1      \\
2.3094   &  1.0 & 2.0 &   0.2  &  8.7  &  0.1   &   5.0  &  5.7   & 1.0  & 1.4  & 4.2& 12.6      \\
2.3864   &  1.0 & 1.8 &   0.2  &  9.0  &  0.1   &   5.0  &  2.8   & 1.0  & 1.6  & 4.1& 11.8      \\
2.3960   &  1.0 & 1.8 &   0.2  &  9.0  &  0.1   &   5.0  &  3.2   & 1.0  & 1.8  & 4.1& 11.9      \\
2.6440   &  1.0 & 1.4 &   0.2  &  5.1  &  0.1   &   5.0  & 11.0   & 1.0  & 1.6  & 3.0& 13.7      \\
2.9000   &  1.0 & 1.5 &   0.2  &  1.3  &  0.1   &   5.0  &  6.6   & 1.0  & 3.2  & 2.6&  9.6      \\
2.9500   &  1.0 & 1.5 &   0.2  &  1.3  &  0.1   &   5.0  &  81    & 1.0  & 3.7  & 2.6& 81.3      \\
\hline
\hline
\end{tabular}
\label{tab:syserrors}
\end{center}
\end{table*}

\begin{table*}[!h]
\begin{center}
\footnotesize
\caption{Summary of all systematic uncertainty sources in Category B. The last column is the summary of the total uncertainties.}
\begin{tabular}{c|cccccccccccc}
\hline
\hline
$\sqrt{s}$~(GeV) & $U_{n\bar{n}}^{B}$ & $U^B_{evt}$ & $U^B_{BDT}$ & $U^B_{fit}$ & $U^B_{model}$ & $U^B_{trig}$  & $U^B_{ISR1}$ & $U^B_{ISR2}$& $U^B_{L}$ & Total (\%)\\
\hline
2.0000	&3.3 & 6.6  & 6.0 & 3.1 & 5.2  & 5.1 & 1.0 & 0.3 & 1.0  & 12.5 \\
2.0500	&2.2 & 5.2  & 6.0 & 6.4 & 9.2  & 5.0 & 1.0 & 0.6 & 1.0  & 14.8\\
2.1000	&1.7 & 5.4  & 6.0 & 0.4 & 9.6  & 4.6 & 1.0 & 1.2 & 1.0  & 13.6\\
2.1250	&1.7 & 5.0  & 6.0 & 1.0 & 9.6  & 4.5 & 1.0 & 1.4 & 1.0  & 13.5\\
2.1500	&1.5 & 6.4  & 3.8 & 1.6 & 9.6  & 4.3 & 1.0 & 1.7 & 1.0  & 13.4\\
2.1750	&1.5 & 6.4  & 3.8 & 1.1 & 6.5  & 4.6 & 1.0 & 1.8 & 1.0  & 11.4\\
2.2000	&1.3 & 6.5  & 3.8 & 0.8 & 10.3 & 4.2 & 1.0 & 2.2 & 1.0  & 13.8\\
2.2320	&1.3 & 6.2  & 3.8 & 2.4 & 9.7  & 4.2 & 1.0 & 2.2 & 1.0  & 13.4\\
2.3090	&1.3 & 6.4  & 3.8 & 1.4 & 8.5  & 3.2 & 1.0 & 1.4 & 1.0  & 12.1\\
2.3860	&1.1 & 3.4  & 3.6 & 1.6 & 10.4 & 2.8 & 1.0 & 1.6 & 1.0  & 12.2\\
2.3960	&1.1 & 3.4  & 3.6 & 0.6 & 7.1  & 3.1 & 1.0 & 1.8 & 1.0  & 9.6\\
2.6440	&0.8 & 9.5  & 3.9 & 1.7 & 7.3  & 2.1 & 1.0 & 1.6 & 1.0  & 13.1\\
2.9000	&0.6 & 9.5  & 4.8 & 1.5 & 7.3  & 1.7 & 1.0 & 3.2 & 1.0  & 13.6 \\
2.9500	&1.0 & 9.7  & 4.8 & 4.3 & 7.3  & 1.4 & 1.0 & 3.7 & 1.0  & 14.4 \\
  \hline
\end{tabular}

\label{tab2:sumsys}
\end{center}
\end{table*}

\begin{table*}[!h]
 \tabcolsep=0.12cm
\begin{center}
\footnotesize
\caption{Summary of all systematic uncertainty sources in Category C. The last column is the summary of the total uncertainties.}
\begin{tabular}{c|ccccccc|c}
  \hline
  \hline
  $\sqrt{s}~\gev$ & $\delta^C_{\rm L}$   & $\delta^C_{\rm sel}$ &  $\delta^C_{\rm fit}$ &  $\delta^C_{\rm model}$ &  $\delta^C_{\rm trg}$  & $\delta^C_{\rm ISR1}$ & $\delta^C_{\rm ISR2}$ & Total (\%)\\
  \hline
2.0000 & 1.0 & 6.6 & 11.8 & 18.8 & 5.3 & $1.0$& 0.3 &23.8  \\
2.0500 & 1.0 & 4.1 & 10.3 & 16.8 & 4.3 & $1.0$& 0.6 &20.6 \\
2.1000 & 1.0 & 4.2 & 9.5  & 15.6 & 4.0 & $1.0$& 1.2 &19.3 \\
2.1250 & 1.0 & 3.2 & 11.1 & 10.0 & 3.9 & $1.0$& 1.4 &15.9  \\
2.1500 & 1.0 & 3.5 & 15.5 & 9.6  & 3.8 & $1.0$& 1.7 &19.1  \\
2.1750 & 1.0 & 3.9 & 11.9 & 8.3  & 4.0 & $1.0$& 1.8 &15.7  \\
2.2000 & 1.0 & 3.9 & 11.4 & 8.3  & 3.8 & $1.0$& 2.2 &15.3  \\
2.2324 & 1.0 & 3.1 & 12.0 & 9.2  & 3.8 & $1.0$& 2.2 &16.1  \\
2.3094 & 1.0 & 3.9 & 6.4  & 7.6  & 2.9 & $1.0$& 1.4 &11.2  \\
2.3864 & 1.0 & 3.4 & 5.5  & 6.3  & 2.5 & $1.0$& 1.6 &9.6   \\
2.3960 & 1.0 & 3.4 & 7.2  & 6.3  & 2.5 & $1.0$& 1.8 &10.7  \\
2.6454 & 1.0 & 3.5 & 16.8 & 9.8  & 1.9 & $1.0$& 1.6 &20.0  \\
2.9000 & 1.0 & 5.5 & 11.9 & 9.3  & 1.7 & $1.0$& 3.2 &16.5  \\
2.9500 & 1.0 & 6.4 & 77.9 & 12.2 & 1.6 & $1.0$& 3.7 &79.2  \\
\hline
\end{tabular}
\label{tab3:sysall}
\end{center}
\end{table*}

\begin{table*}[h]
\scriptsize
   \tabcolsep=0.15cm
\renewcommand{\arraystretch}{1.4}
\begin{center}
\caption{ The form factor ratios $R_{em} = |G_E|/|G_M|$, electric $|G_E|$ and magnetic $|G_M|$ form factors.
The results are obtained by fitting angular distributions of the antineutron w/o systematic uncertainties at each bin.
The center shifts of results w/o considering systematic uncertainties per $\cos\theta_{\bar{n}}$ during the fitting
is also taken as a systematic uncertainty added to final results.
The label "stat. only" represents results obtained from the fitting to data of antineutron angular distributions with statistical errors. The label "stat. + sys."  represents results obtained from the fitting to data of antineutron angular distributions with both statistical errors and systematic errors.}
\begin{tabular}{c|cccc}
	\hline
	\hline
\multirow{3}{*}{$\sqrt{s}$ (GeV)}  & \multicolumn{4}{c}{$R_{em}$}\\
\cline{2-5}
  & \multirow{2}{*}{stat. only} &  \multirow{2}{*}{stat. + sys.}   &  center    & total sys. \\
  &  &  & shift & uncertainty \\
	\hline
2.0000,2.0500               & 0.74$\pm$0.69 & 0.93$\pm$0.78 & 0.19  & 0.40 \\
2.1250,2.1500               & 1.32$\pm$0.37 & 1.29$\pm$0.45 & 0.02  & 0.25\\
2.1750,2.2000,2.2324, 2.3094 & 1.53$\pm$0.64 & 1.51$\pm$0.67 & 0.03  & 0.19\\
2.3864,2.3960               & 0.89$\pm$0.34 & 0.88$\pm$0.38 & 0.02  & 0.16\\
2.6454,2.9500               & 0.90$\pm$0.77 & 0.56$\pm$0.99 & 0.34  & 0.70\\
\hline
\multirow{3}{*}{$\sqrt{s}$(\gev)}  & \multicolumn{4}{c}{$|G_M|$ $(10^{-2})$}\\
\cline{2-5}
   & \multirow{2}{*}{stat. only} &  \multirow{2}{*}{stat. + sys.}   &  center    & total sys.\\
  &  &  & shift & uncertainty \\
\hline
2.0000,2.0500               &20.3$\pm$5.0 & 18.6$\pm$5.6 & 1.8 & 3.1 \\
2.1250,2.1500               & 8.7$\pm$1.2 & 8.7$\pm$1.5  & 0.0 & 0.8\\
2.1750,2.2000,2.2324, 2.3094 & 6.6$\pm$1.5 & 6.5$\pm$1.6 & 0.1 & 0.4\\
2.3864,2.3960               & 8.4$\pm$0.9 & 8.3$\pm$0.9  & 0.0 & 0.4\\
2.6454,2.9500               & 4.2$\pm$0.8 & 4.4$\pm$0.9  & 0.2 & 0.3\\
	\hline
\hline
\multirow{3}{*}{$\sqrt{s}$(\gev)}  & \multicolumn{4}{c}{$|G_E|$ $(10^{-2})$}\\
\cline{2-5}
   & correlation &  \multirow{2}{*}{stat. only}   &  center    & total \\
  &  coefficients &  & shift & uncertainty \\
\hline
2.0000,2.0500               &-0.984 &17.2$\pm$8.3 & - & 4.7\\
2.1250,2.1500               &-0.990 &11.2$\pm$1.7 & - & 1.1\\
2.1750,2.2000,2.2324, 2.3094 &-0.992 & 9.8$\pm$1.9 & - & 0.6\\
2.3864,2.3960               &-0.979 & 7.3$\pm$2.1 & - & 1.0\\
2.6454,2.9500               &-0.968 & 2.5$\pm$2.9 & - & 2.9\\
	\hline
\end{tabular}
\label{tab:angresults3}
\end{center}
\end{table*}

\clearpage
\section{DATA}
\hypertarget{item:h}{}

\begin{table*}[h]\footnotesize
    \centering
    \caption{Differential cross-sections for category A, in unit of pb.}
    \begin{tabular}{|c|c|c|c|c|c|c|c|}
    \hline
 $\sqrt{s}$ GeV & bin1 & bin2 & bin3 & bin4 & bin5 & bin6 & bin7\\
    \hline
2.0000  &$18.9_{- 9.7}^{+12.9}\pm3.0$&$35.4_{-14.3}^{+18.0}\pm3.0$&$30.2_{-11.0}^{+14.0}\pm3.0$&$37.3_{-17.1}^{+21.4}\pm3.0$&$29.9_{-13.3}^{+16.4}\pm3.0$&$30.9_{-12.5}^{+15.9}\pm3.0$&$26.4_{-11.4}^{+14.9}\pm3.0$\\
2.0500  &$ 0.0_{- 0.0}^{+30.3}\pm0.0$&$ 0.0_{- 0.0}^{+ 7.2}\pm0.0$&$42.9_{-21.2}^{+30.5}\pm0.0$&$16.7_{-21.1}^{+37.0}\pm0.0$&$12.7_{-19.5}^{+25.3}\pm0.0$&$100.0_{-35.0}^{+44.5}\pm0.0$&$25.2_{-16.9}^{+24.9}\pm0.0$\\
2.1250  &$ 9.4_{- 2.0}^{+ 2.3}\pm1.3$&$16.7_{- 2.8}^{+ 3.2}\pm1.3$&$16.6_{- 3.2}^{+ 3.5}\pm1.3$&$17.8_{- 3.6}^{+ 4.0}\pm1.3$&$14.3_{- 3.0}^{+ 3.4}\pm1.3$&$12.7_{- 2.6}^{+ 2.8}\pm1.3$&$12.4_{- 2.3}^{+ 2.6}\pm1.3$\\
2.1500  &$15.2_{-11.5}^{+22.4}\pm2.5$&$16.1_{-16.1}^{+26.9}\pm2.5$&$ 0.0_{- 0.0}^{+ 9.8}\pm2.5$&$29.0_{-26.6}^{+49.8}\pm2.5$&$ 0.0_{- 0.0}^{+10.5}\pm2.5$&$51.4_{-26.3}^{+38.1}\pm2.5$&$ 0.0_{- 0.0}^{+ 7.8}\pm2.5$\\
2.1750  &$ 9.1_{- 5.5}^{+ 8.8}\pm1.5$&$13.9_{- 7.2}^{+10.5}\pm1.5$&$ 0.0_{- 0.0}^{+ 5.5}\pm1.5$&$30.8_{-16.0}^{+20.2}\pm1.5$&$ 8.7_{- 7.2}^{+11.4}\pm1.5$&$12.8_{- 6.9}^{+10.0}\pm1.5$&$ 0.8_{- 0.0}^{+ 7.1}\pm1.5$\\
2.2200  &$ 2.8_{- 0.0}^{+ 5.4}\pm0.4$&$ 6.9_{- 5.3}^{+ 8.4}\pm0.4$&$ 0.0_{- 0.0}^{+ 2.7}\pm0.4$&$18.9_{- 9.3}^{+12.3}\pm0.4$&$ 7.6_{- 4.9}^{+ 7.5}\pm0.4$&$ 0.9_{- 0.0}^{+ 8.2}\pm0.4$&$ 2.9_{- 2.7}^{+ 5.0}\pm0.4$\\
2.2324  &$ 3.6_{- 2.7}^{+ 5.3}\pm0.5$&$ 3.3_{- 2.2}^{+ 7.4}\pm0.5$&$13.8_{- 9.0}^{+13.0}\pm0.5$&$ 1.5_{- 1.1}^{+ 9.1}\pm0.5$&$16.1_{- 8.2}^{+11.4}\pm0.5$&$ 0.0_{- 0.0}^{+ 3.9}\pm0.5$&$ 9.5_{- 5.9}^{+ 8.5}\pm0.5$\\
2.3094  &$ 3.7_{- 2.7}^{+ 4.4}\pm0.5$&$14.0_{- 5.2}^{+ 6.8}\pm0.5$&$17.5_{- 7.3}^{+ 9.5}\pm0.5$&$ 8.2_{- 6.4}^{+ 8.7}\pm0.5$&$13.3_{- 5.8}^{+ 7.6}\pm0.5$&$11.1_{- 6.3}^{+ 8.2}\pm0.5$&$ 4.5_{- 4.4}^{+ 6.0}\pm0.5$\\
2.3864  &$ 8.6_{- 4.0}^{+ 5.3}\pm1.1$&$11.3_{- 4.8}^{+ 6.4}\pm1.1$&$ 3.2_{- 3.1}^{+ 4.7}\pm1.1$&$11.5_{- 5.8}^{+ 7.4}\pm1.1$&$ 3.6_{- 3.5}^{+ 5.4}\pm1.1$&$15.2_{- 6.0}^{+ 7.7}\pm1.1$&$ 7.4_{- 4.1}^{+ 5.4}\pm1.1$\\
2.3960  &$10.2_{- 2.4}^{+ 2.8}\pm1.3$&$10.6_{- 2.9}^{+ 3.4}\pm1.3$&$ 6.8_{- 2.4}^{+ 3.0}\pm1.3$&$14.7_{- 4.1}^{+ 4.8}\pm1.3$&$ 7.0_{- 2.8}^{+ 3.4}\pm1.3$&$ 9.6_{- 2.7}^{+ 3.2}\pm1.3$&$ 9.8_{- 2.5}^{+ 2.9}\pm1.3$\\
2.6450  &$ 3.2_{- 1.5}^{+ 2.1}\pm0.5$&$ 2.6_{- 1.4}^{+ 2.3}\pm0.5$&$ 0.0_{- 0.0}^{+ 0.6}\pm0.5$&$ 9.9_{- 3.7}^{+ 4.8}\pm0.5$&$ 4.3_{- 2.0}^{+ 2.8}\pm0.5$&$ 3.0_{- 1.9}^{+ 2.8}\pm0.5$&$ 3.9_{- 1.7}^{+ 2.3}\pm0.5$\\
2.9500  &$ 0.0_{- 0.0}^{+ 1.8}\pm0.0$&$ 0.0_{- 0.0}^{+ 1.8}\pm0.0$&$ 0.0_{- 0.0}^{+ 2.5}\pm0.0$&$ 0.0_{- 0.0}^{+ 2.8}\pm0.0$&$ 2.4_{- 2.4}^{+ 6.5}\pm0.0$&$ 4.2_{- 4.2}^{+ 6.7}\pm0.0$&$ 0.0_{- 0.0}^{+ 1.6}\pm0.0$\\
\hline
\hline
\end{tabular}
\label{tab:diffxsA}
%\end{table*}
%\begin{table*}[h]\footnotesize
    \centering
    \caption{Differential cross-sections for category B, in unit of pb.}
    \begin{tabular}{|c|c|c|c|c|c|c|c|}
    \hline
 $\sqrt{s}$ GeV & bin1 & bin2 & bin3 & bin4 & bin5 & bin6 & bin7\\
    \hline
2.0000  &$24.8_{-18.8}^{+46.0}\pm3.1$&$75.3_{-39.2}^{+68.2}\pm3.1$&$45.8_{-29.6}^{+60.5}\pm3.1$&$38.7_{-27.8}^{+63.9}\pm3.1$&$55.5_{-35.8}^{+73.3}\pm3.1$&$ 0.0_{- 0.0}^{+26.4}\pm3.1$&$55.1_{-32.4}^{+61.3}\pm3.1$\\
2.0500  &$ 0.0_{- 0.0}^{+50.1}\pm0.0$&$55.6_{-38.3}^{+83.1}\pm0.0$&$ 0.0_{- 0.0}^{+34.7}\pm0.0$&$ 0.0_{- 0.0}^{+34.2}\pm0.0$&$ 0.0_{- 0.0}^{+35.7}\pm0.0$&$ 0.0_{- 0.0}^{+33.2}\pm0.0$&$ 0.0_{- 0.0}^{+46.7}\pm0.0$\\
2.1250  &$10.9_{- 2.5}^{+ 3.1}\pm1.5$&$13.6_{- 2.3}^{+ 2.8}\pm1.5$&$12.7_{- 2.2}^{+ 2.7}\pm1.5$&$11.6_{- 2.1}^{+ 2.6}\pm1.5$&$10.1_{- 2.0}^{+ 2.5}\pm1.5$&$ 9.3_{- 2.0}^{+ 2.4}\pm1.5$&$14.5_{- 2.8}^{+ 3.4}\pm1.5$\\
2.1500  &$20.9_{-15.9}^{+38.8}\pm3.0$&$38.6_{-21.3}^{+38.5}\pm3.0$&$35.6_{-19.3}^{+34.6}\pm3.0$&$ 0.0_{- 0.0}^{+13.0}\pm3.0$&$16.2_{-12.3}^{+30.0}\pm3.0$&$11.3_{- 9.6}^{+28.3}\pm3.0$&$42.4_{-24.6}^{+45.8}\pm3.0$\\
2.1750  &$14.0_{- 7.1}^{+12.1}\pm1.7$&$ 9.4_{- 5.2}^{+ 9.4}\pm1.7$&$13.0_{- 6.1}^{+10.1}\pm1.7$&$10.2_{- 5.4}^{+ 9.5}\pm1.7$&$ 6.4_{- 4.1}^{+ 8.5}\pm1.7$&$ 6.8_{- 4.4}^{+ 8.9}\pm1.7$&$10.5_{- 6.4}^{+12.5}\pm1.7$\\
2.2200  &$ 8.3_{- 4.5}^{+ 8.1}\pm1.2$&$ 4.1_{- 2.7}^{+ 5.8}\pm1.2$&$ 4.7_{- 3.0}^{+ 5.9}\pm1.2$&$ 2.9_{- 2.2}^{+ 5.5}\pm1.2$&$ 7.5_{- 3.9}^{+ 6.8}\pm1.2$&$ 3.6_{- 2.6}^{+ 5.7}\pm1.2$&$20.3_{- 7.3}^{+10.5}\pm1.2$\\
2.2324  &$ 0.0_{- 0.0}^{+ 3.2}\pm0.0$&$ 0.5_{- 0.5}^{+ 4.9}\pm0.0$&$ 9.0_{- 4.5}^{+ 7.7}\pm0.0$&$16.4_{- 6.2}^{+ 9.2}\pm0.0$&$10.5_{- 5.0}^{+ 8.2}\pm0.0$&$11.1_{- 5.1}^{+ 8.3}\pm0.0$&$15.1_{- 6.7}^{+10.6}\pm0.0$\\
2.3094  &$ 4.4_{- 2.4}^{+ 4.4}\pm0.6$&$ 6.6_{- 2.9}^{+ 4.7}\pm0.6$&$ 7.2_{- 3.1}^{+ 4.8}\pm0.6$&$11.1_{- 3.9}^{+ 5.5}\pm0.6$&$13.2_{- 4.2}^{+ 5.8}\pm0.6$&$ 5.2_{- 2.5}^{+ 4.3}\pm0.6$&$ 8.7_{- 3.7}^{+ 5.7}\pm0.6$\\
2.3864  &$ 4.2_{- 2.1}^{+ 3.5}\pm0.5$&$10.0_{- 3.0}^{+ 4.1}\pm0.5$&$10.0_{- 3.1}^{+ 4.2}\pm0.5$&$ 6.2_{- 2.4}^{+ 3.6}\pm0.5$&$ 6.0_{- 2.3}^{+ 3.4}\pm0.5$&$12.7_{- 3.4}^{+ 4.5}\pm0.5$&$ 7.1_{- 2.7}^{+ 4.0}\pm0.5$\\
2.3960  &$ 9.0_{- 1.8}^{+ 2.3}\pm0.8$&$ 7.9_{- 1.6}^{+ 2.0}\pm0.8$&$ 7.6_{- 1.6}^{+ 1.9}\pm0.8$&$ 8.6_{- 1.7}^{+ 2.1}\pm0.8$&$ 6.9_{- 1.5}^{+ 1.8}\pm0.8$&$ 8.7_{- 1.7}^{+ 2.0}\pm0.8$&$10.2_{- 2.0}^{+ 2.4}\pm0.8$\\
2.6450  &$ 2.6_{- 0.8}^{+ 1.1}\pm0.4$&$ 1.4_{- 0.6}^{+ 0.9}\pm0.4$&$ 2.2_{- 0.7}^{+ 1.0}\pm0.4$&$ 2.4_{- 0.8}^{+ 1.1}\pm0.4$&$ 1.4_{- 0.6}^{+ 0.9}\pm0.4$&$ 2.2_{- 0.7}^{+ 1.0}\pm0.4$&$ 2.2_{- 0.8}^{+ 1.1}\pm0.4$\\
\hline
\hline
\end{tabular}
\label{tab:diffxsB}
%\end{table*}
%\begin{table*}[h]\footnotesize
    \centering
    \caption{Differential cross-sections for category C, in unit of pb.}
    \begin{tabular}{|c|c|c|c|c|c|c|c|}
    \hline
 $\sqrt{s}$ GeV & bin1 & bin2 & bin3 & bin4 & bin5 & bin6 & bin7\\
    \hline
2.0000  &$93.8_{-45.5}^{+75.5}\pm24.4$&$22.1_{-16.4}^{+38.5}\pm24.4$&$ 0.0_{- 0.0}^{+14.2}\pm24.4$&$14.8_{-13.0}^{+41.2}\pm24.4$&$23.1_{-17.2}^{+40.2}\pm24.4$&$72.2_{-31.2}^{+48.8}\pm24.4$&$92.0_{-35.6}^{+53.1}\pm24.4$\\
2.0500  &$54.3_{-34.4}^{+69.0}\pm12.1$&$19.7_{-16.3}^{+45.2}\pm12.1$&$ 0.0_{- 0.0}^{+19.9}\pm12.1$&$73.4_{-38.8}^{+68.1}\pm12.1$&$ 0.0_{- 0.0}^{+19.2}\pm12.1$&$ 8.0_{- 7.8}^{+40.8}\pm12.1$&$48.4_{-30.1}^{+59.4}\pm12.1$\\
2.1250  &$10.3_{- 1.9}^{+ 2.3}\pm1.7$&$ 9.0_{- 1.6}^{+ 1.9}\pm1.7$&$ 9.2_{- 1.6}^{+ 1.8}\pm1.7$&$ 7.8_{- 1.6}^{+ 1.9}\pm1.7$&$ 8.4_{- 1.5}^{+ 1.8}\pm1.7$&$12.4_{- 1.9}^{+ 2.2}\pm1.7$&$ 8.6_{- 1.6}^{+ 2.0}\pm1.7$\\
2.1500  &$ 9.6_{- 8.2}^{+24.0}\pm1.9$&$ 0.0_{- 0.0}^{+ 8.9}\pm1.9$&$27.3_{-14.8}^{+26.6}\pm1.9$&$19.5_{-12.6}^{+25.7}\pm1.9$&$ 7.3_{- 6.4}^{+20.3}\pm1.9$&$ 0.0_{- 0.0}^{+ 8.4}\pm1.9$&$ 0.0_{- 0.0}^{+ 8.9}\pm1.9$\\
2.1750  &$ 7.4_{- 3.9}^{+ 7.0}\pm1.2$&$ 6.5_{- 3.4}^{+ 6.0}\pm1.2$&$ 2.1_{- 1.8}^{+ 4.9}\pm1.2$&$12.3_{- 5.4}^{+ 8.5}\pm1.2$&$ 4.5_{- 2.9}^{+ 5.8}\pm1.2$&$12.0_{- 5.1}^{+ 8.0}\pm1.2$&$ 4.9_{- 3.1}^{+ 6.3}\pm1.2$\\
2.2200  &$ 6.1_{- 3.1}^{+ 5.3}\pm1.0$&$ 7.9_{- 3.3}^{+ 5.1}\pm1.0$&$10.9_{- 4.1}^{+ 6.1}\pm1.0$&$ 1.1_{- 1.0}^{+ 3.3}\pm1.0$&$ 2.3_{- 1.7}^{+ 3.9}\pm1.0$&$15.2_{- 4.6}^{+ 6.3}\pm1.0$&$ 5.9_{- 2.9}^{+ 4.8}\pm1.0$\\
2.2324  &$ 3.1_{- 2.1}^{+ 4.4}\pm0.5$&$ 6.3_{- 3.1}^{+ 5.1}\pm0.5$&$ 5.0_{- 2.7}^{+ 4.7}\pm0.5$&$ 8.2_{- 3.8}^{+ 6.2}\pm0.5$&$ 1.3_{- 1.2}^{+ 3.6}\pm0.5$&$10.6_{- 4.0}^{+ 5.8}\pm0.5$&$ 5.6_{- 2.9}^{+ 5.0}\pm0.5$\\
2.3094  &$12.4_{- 3.5}^{+ 4.6}\pm1.5$&$ 7.4_{- 2.6}^{+ 3.7}\pm1.5$&$ 0.6_{- 0.5}^{+ 2.1}\pm1.5$&$ 3.5_{- 1.9}^{+ 3.3}\pm1.5$&$ 7.7_{- 2.7}^{+ 3.9}\pm1.5$&$11.4_{- 3.2}^{+ 4.2}\pm1.5$&$ 1.6_{- 1.1}^{+ 2.4}\pm1.5$\\
2.3864  &$ 9.2_{- 2.8}^{+ 3.8}\pm1.0$&$ 3.6_{- 1.6}^{+ 2.6}\pm1.0$&$ 9.7_{- 2.8}^{+ 3.8}\pm1.0$&$ 5.4_{- 2.1}^{+ 3.1}\pm1.0$&$ 3.5_{- 1.6}^{+ 2.6}\pm1.0$&$ 4.6_{- 1.8}^{+ 2.6}\pm1.0$&$ 6.7_{- 2.2}^{+ 3.2}\pm1.0$\\
2.3960  &$ 6.2_{- 1.4}^{+ 1.7}\pm0.7$&$ 4.4_{- 1.1}^{+ 1.4}\pm0.7$&$ 7.9_{- 1.5}^{+ 1.8}\pm0.7$&$ 6.0_{- 1.3}^{+ 1.6}\pm0.7$&$ 7.2_{- 1.4}^{+ 1.7}\pm0.7$&$ 7.9_{- 1.4}^{+ 1.7}\pm0.7$&$ 8.4_{- 1.5}^{+ 1.8}\pm0.7$\\
2.6450  &$ 1.1_{- 0.5}^{+ 0.7}\pm0.3$&$ 1.0_{- 0.4}^{+ 0.6}\pm0.3$&$ 2.1_{- 0.6}^{+ 0.8}\pm0.3$&$ 0.2_{- 0.2}^{+ 0.5}\pm0.3$&$ 0.3_{- 0.2}^{+ 0.5}\pm0.3$&$ 2.0_{- 0.6}^{+ 0.8}\pm0.3$&$ 0.8_{- 0.4}^{+ 0.6}\pm0.3$\\
2.9500  &$ 0.0_{- 0.0}^{+ 0.6}\pm0.0$&$ 0.0_{- 0.0}^{+ 0.6}\pm0.0$&$ 0.0_{- 0.0}^{+ 0.6}\pm0.0$&$ 2.5_{- 1.3}^{+ 2.2}\pm0.0$&$ 1.0_{- 0.8}^{+ 1.7}\pm0.0$&$ 1.9_{- 1.0}^{+ 1.7}\pm0.0$&$ 0.0_{- 0.0}^{+ 0.6}\pm0.0$\\
\hline
\hline
\end{tabular}
\label{tab:diffxsC}
\end{table*}

\clearpage

\begin{table}[h!]
\tiny
   \tabcolsep=0.05cm
\renewcommand{\arraystretch}{1.5}
  \begin{center}
  \caption{DATA: $N_{signal}$ @CatA}
\begin{tabular}{c | ccccccccccccccccccccc | c}
$\sqrt{s}$ & BIN1 & errL & errH & BIN2 & errL & errH & BIN3 & errL & errH & BIN4 & errL & errH & BIN5 & errL & errH & BIN6 & errL & errH & BIN7 & errL & errH & tot       \\
\hline
\hline
2.0000 & 4.3 &  2.2  & 2.9 &  6.6 &  2.7 &  3.4 &  7.0 &  2.5 &  3.3 &  6.3 &  2.9 &  3.6 &  6.5 &  2.9 &  3.6 &  6.4 &  2.6 &  3.3 &  5.7 &  2.5 &  3.2 &  47.1        \\
2.0500 & 0 &  0 &  2.1 &  0  & 0 &  0.5 &  3.2  & 1.6 &  2.3 &  1.0 &  0  & 2.1  & 0.7 &  0  & 1.4  & 7.3 &  2.5  & 3.2 &  2.2  & 1.5 &  2.2  & 14.4    \\
2.1250 & 22.8 &  5.0  & 5.7 &  37.6 &  6.4 &  7.1  & 32.4  & 6.2 &  6.9 &  34.2 &  6.9  & 7.7  & 29.2  & 6.2  & 6.9 &  28.6 &  5.8  & 6.4  & 30.2 &  5.7  & 6.5 &  237.8    \\
2.1500 & 0.9 &   0.7  &  1.4  &  1.0  &  1.0 &   1.6  &  0 &   0 &   0.5  &  1.3 &   0 &   2.2 &   0  &  0 &   0.5  &  3.3  &  1.7 &   2.4 &   0 &    0 &   0.5  &  7.4     \\
2.1750 & 1.9  &  1.2  &  1.9   & 2.8  &  1.5  &  2.1 &   0  &  0  &  1.0  &  5.3  &  2.8 &   3.5 &   1.4 &   1.2 &   1.9  &  2.7  &  1.5  &  2.2   & 0.2  &  0  &  1.5 &   16.2    \\
2.2000 & 0.8  &  0 &   1.5 &   1.6   & 1.2  &  1.9  &  0  &  0  &  0.6  &  4.2 &   2.0  &  2.7  &  2.1 &   1.3  &  2.0 &   0.2 &   0  &  2.1   & 0.8 &   0.8  &  1.4  &  10.5        \\
2.2324 & 1.0  &  0.7  &  1.4 &   0.8  &  0  &  1.7  &  2.6  &  1.7  &  2.5 &   0.3  &  0 &   1.7  &  3.4  &  1.8 &   2.4  &  0 &   0  &  0.8  &  2.7 &   1.7  &  2.4  &  11.8    \\
2.3094 & 1.5 &   1.1  &  1.8 &   5.9  &  2.2 &   2.9 &   6.0  &  2.5  &  3.3 &   2.7  &  2.1 &   2.9  &  5.2  &  2.3  &  3.0  &  4.5  &  2.6  &  3.3 &   2.2 &   2.1  &  2.9 &   29.5  \\
2.3864 & 4.6 &   2.1  &  2.8  &  5.0  &  2.1  &  2.8  &  1.5  &  1.4  &  2.2 &   5.2  &  2.6  &  3.4  &  1.4  &  2.1 &   2.2  &  7.3  &  2.9  &  3.7 &   4.1  &  2.2  &  3 &   33.7  \\
2.3960 & 16.5  &  3.8  &  4.5 &   14.1  &  3.9  &  4.6 &   9.4  &  3.4  &  4.2 &   17.5  &  5.0  &  5.7  &  8.7  &  3.5 &   4.2 &   13.9  &  3.9  &  4.6 &   15.3  &  3.9 &   4.6  &  111.9   \\
2.6454 & 3.5  &  1.6 &   2.3  &  2.0  &  1.1 &   1.8  &  0  &  0 &   0.5  &  6.1 &   2.3 &   3.0 &   3.5  &  1.6  &  2.3  &  2.5 &   1.6 &   2.3  &  4.0  &  1.7 &   2.4 &   25.1      \\
2.9500 & 0 &   0  &  0.5  &  0  &  0 &   0.5  &  0  &  0 &   0.5  &  0  &  0  &  0.5  &  0.5 &   0  &  1.4  &  1.1  &  1.1  &  1.8  &  0  &  0  &  0.5 &   0
\end{tabular}
  \end{center}
%\end{table}
%\clearpage
%\begin{table}
\tiny
   \tabcolsep=0.05cm
\renewcommand{\arraystretch}{1.5}
  \begin{center}
\caption{DATA: $N_{signal}$ @CatB}
\begin{tabular}{c | ccccccccccccccccccccc | c}
$\sqrt{s}$ & BIN1 & errL & errH & BIN2 & errL & errH & BIN3 & errL & errH & BIN4 & errL & errH & BIN5 & errL & errH & BIN6 & errL & errH & BIN7 & errL & errH & tot        \\
\hline
\hline
2.0000  & 1.3   &  1.0  &   2.4   &  3.3   &  1.7  &   3.0   &  2.0    & 1.3   &  2.6   &  1.5  &   1.1  &   2.5  &   2.0  &   1.3   & 2.6  &   0   &  0   &  1.0  &   2.5  &   1.5   &  2.8   &  12.8   \\
2.0500  & 0   &  0    & 1.0  &   1.7  &   1.2    & 2.5    & 0    & 0   &  1.0  &   0  &   0  &   1.0  &   0   &  0  &   1.0   &  0   &  0   &  1.0   &  0   &  0   &  1.0  &   1.7    \\
2.1250  & 19.3   &  4.4    & 5.5   &  33.6  &   5.8  &   6.9   &  31.4   &  5.6  &   6.7  &   28.8  &   5.3  &   6.4  &   24.6   &  4.9  &   6.0   &  22.2  &   4.7   &  5.8  &   25.9  &   5.1   &  6.2  &   205.0         \\
2.1500  & 1.3   &  1.0  &   2.4   &  2.9   &  1.6  &   2.9  &   3.0    & 1.6  &   2.9  &   0   &  0  &   1.0   &  1.3 &    1.0   &  2.4   &  0.9  &   0.8  &   2.3   &  2.6   &  1.5   &  2.8  &   12.2  \\
2.1750  & 3.5    & 1.8 &    3.0   &  2.9   &  1.6  &   2.9  &   4.1 &    1.9  &   3.2   &  3.2  &   1.7   &  3.0   &  2.0  &   1.3   &  2.6   &  2.0  &   1.3  &   2.6   &  2.3   &  1.4  &   2.7  &   23.5   \\
2.2000  & 3.0  &   1.6  &   2.9  &   1.8   &  1.2  &   2.6  &   2.1   &  1.3   &  2.7   &  1.3   &  1.0   &  2.41   &  3.3   &  1.7   &  3.0  &   1.6    & 1.1  &   2.5   &  7.4  &   2.7   &  3.8  &   23.5   \\
2.2000  & 0   &  0   &  1.0   &  0.2   &  0.2    & 2.0   &  3.6    & 1.8  &   3.1  &   6.6  &   2.5  &   3.7  &   4.1  &   1.9   &  3.2   &  4.3  &   2.0   &  3.2   &  4.7  &   2.1   &  3.3  &   23.5    \\
2.3094  & 2.9  &   1.6   &  2.9  &   4.7  &   2.1   &  3.3  &   5.1   &  2.2   &  3.4   &  7.9   &  2.8  &   3.9   &  9.6  &   3.0   &  4.2   &  3.8  &   1.9  &   3.12   &  5.2  &   2.21   &  3.4   &  42.1    \\
2.3864  & 3.7   &  1.8  &   3.1  &   10.6   &  3.2   &  4.4  &   10.3  &   3.2   &  4.3  &   6.1   &  2.4  &   3.6    & 6.5  &   2.5    & 3.7   &  13.3   &  3.6  &   4.7  &   6.5  &   2.5  &   3.7   &  60.7      \\
2.3960  & 23.7   &  4.8  &   5.9   &  23.7  &   4.8  &   5.9   &  22.7   &  4.7  &   5.8  &   25.8   &  5.1  &   6.1   &  21.7   &  4.6   &  5.7  &   26.7   &  5.1  &   6.2   &  26.7   &  5.1   &  6.2  &   195.0         \\
2.6454  & 9.5   &  3.0  &   4.2  &   5.9  &   2.4   &  3.6   &  8.6  &   2.9   &  4.1   &  9.3   &  3.0  &   4.2   &  5.8   &  2.3  &   3.5  &   9.0  &   2.9  &   4.1  &   7.9  &   2.8  &   3.9    & 65.5   \\
2.9500  & 0   &  0   &  1.0   &  1.0  &   0.8  &   2.3  &   0   &  0  &   1.0  &   0   &  0   &  1.0  &   0   &  0  &   1.0   &  1.0  &   0.8  &   2.3    & 4.1   &  1.9 &    3.2  &   6.1         \\
\end{tabular}
  \end{center}
%\end{table}
%\clearpage
%\begin{table}
\tiny
   \tabcolsep=0.05cm
\renewcommand{\arraystretch}{1.5}
  \begin{center}
\caption{DATA: $N_{signal}$ @CatC}
\begin{tabular}{c | ccccccccccccccccccccc | c}
$\sqrt{s}$ & BIN1 & errL & errH & BIN2 & errL & errH & BIN3 & errL & errH & BIN4 & errL & errH & BIN5 & errL & errH & BIN6 & errL & errH & BIN7 & errL & errH & tot        \\
\hline
\hline
2.0000  & 3.9  &   1.9   &  3.1  &   1.4   &  1.0   &  2.4   &  0   &  0   &  1.0   &  0.8   &  0.7   &  2.2   &  1.4   &  1.0   &  2.4 &    5.0  &   2.2   &  3.4  &   6.3  &   2.4  &   3.6   &  18.8       \\
2.0500  & 2.1  &   1.3   &  2.7  &   1.0  &   0.8  &   2.3  &   0   &  0   &  1.0  &   3.2   &  1.7  &   3.0  &   0  &   0   &  1.0   &  0.4   &  0.4   &  2.1  &   2.2   &  1.4   &  2.7   &  8.9    \\
2.1250  & 29.7  &   5.4  &   6.5   &  32.7   &  5.7  &   6.8   &  34.6  &   5.9   &  6.9  &   24.5   &  4.9   &  6.0   &  31.4   &  5.6   &  6.7   &  43.7  &   6.6   &  7.7   &  27.7    & 5.2   &  6.3   &  254.0   \\
2.1500  & 0.9  &   0.8  &   2.3  &   0   &  0  &   1.0   &  3.0 &    1.6   &  2.9   &  2.0   &  1.3   &  2.6  &   0.8   &  0.7  &   2.2  &   0  &   0  &   1.0   &  0  &   0  &   1.0   &  6.7   \\
2.1750  & 3.1  &   1.7   &  2.9   &  3.2  &   1.7   &  3.0   &  1.0   &  0.8   &  2.3  &   4.9   &  2.1   &  3.4   &  2.1  &   1.3   &  2.7   &  5.1  &   2.2  &   3.4  &   2.1  &   1.3  &   2.7    & 24.6       \\
2.2000  & 3.5   &  1.8  &   3.0  &   5.3  &   2.2  &   3.4  &   6.6   &  2.5  &   3.7  &   0.7  &   0.6  &   2.2   &  1.5   &  1.1    & 2.5   &  10.5   &  3.2   &  4.3   &  3.9   &  1.9  &   3.1   &  35.5        \\
2.2324  & 1.8   &  1.2   &  2.6  &   3.9  &   1.9   &  3.1   &  3.1  &   1.7   &  2.9   &  4.3   &  2.0 &    3.2  &   0.8   &  0.7  &   2.2  &   6.8  &   2.5  &   3.7   &  3.4   &  1.8  &   3.0  &   25.9      \\
2.3094  & 12.4  &   3.5  &   4.6   &  7.8   &  2.7   &  3.9   &  0.6  &   0.6  &   2.1   &  3.1   &  1.7  &   2.9   &  7.6  &   2.7    & 3.9  &   12.7   &  3.5  &   4.7   &  1.7    & 1.2  &   2.5   &  58.3      \\
2.3864  & 10.5  &   3.2  &   4.3   &  4.5   &  2.0  &   3.3   &  11.6  &   3.4   &  4.5  &   6.3  &   2.4   &  3.6   &  4.3  &   2.0   &  3.23   &  6.4  &   2.5  &   3.7  &   8.6  &   2.9  &   4.1  &   62.7      \\
2.3960  & 20  &   4.4   &  5.6   &  17.0  &   4.1  &   5.2 &    28.3   &  5.3  &   6.4   &  21.0   &  4.6  &   5.7    & 25.8   &  5.1   &  6.1  &   31.8  &   5.6  &   6.7  &   32.2  &   5.6  &   6.7  &   196.0     \\
2.6454  & 5.2   &  2.2  &   3.4  &   5.9  &   2.4   &  3.6   &  12.2   &  3.4   &  4.6  &   0.9   &  0.8   &  2.3  &   1.7   &  1.2  &   2.5  &  10.5   &  3.2  &   4.3  &   4.2  &   2.0  &   3.2  &   45.8    \\
2.9500  & 0  &   0  &   1.0   &  0   &  0  &   1.0   &  0  &   0  &   1.0   &  3.4  &   1.8   &  3.0   &  1.5   &  1.1  &   2.5  &   3.2   &  1.7   &  3.0  &   0  &   0   &  1.0   &  8.1
\end{tabular}
  \end{center}
\end{table}

\clearpage

\begin{table}
\tiny
   \tabcolsep=0.15cm
\renewcommand{\arraystretch}{1.5}
  \begin{center}
\begin{tabular}{c | ccccccc}
$\sqrt{s}$ & BIN1 & BIN2 & BIN3 &  BIN4 &  BIN5 &  BIN6 &  BIN7   \\
\hline
\hline
2.0000  & 0.0238  &     0.0232  &     0.0232   &    0.0230  &     0.0234  &     0.0239   &    0.0246      \\
2.0500  &0.0250  &  0.0230   &    0.0227   &    0.0209  &     0.0208   &    0.0232   &    0.0242     \\
2.1250  &0.0224   &    0.0214  &     0.0195  &     0.0183  &     0.0200   & 0.021   &    0.0221     \\
2.1500  &0.0217  &     0.0194  &     0.0192   &    0.0177  &     0.0183  &    0.0220   & 0.0217     \\
2.1750  &0.0206   &    0.0188 &      0.0169   &    0.0167  &     0.0170  &  0.0195  &     0.0205     \\
2.2000  &0.0194   &    0.0176  &     0.0163  &     0.0162  &     0.0180  &  0.0187  &     0.0207     \\
2.2324  &0.0209   &    0.0188  &     0.0160  &     0.0171   &    0.0180  &  0.0183  &     0.0209     \\
2.3904  &0.0226   &   0.0219   &    0.0186  &     0.0179   &    0.0201  &     0.0210  &     0.0232     \\
2.3864  &0.0254   &    0.0234  &     0.0223  &     0.0230  &     0.0224  &     0.0248   &    0.0264     \\
2.3960  &0.0271   &    0.0242  &     0.0224  &     0.0207   &    0.0224 &      0.0244  &     0.0255     \\
2.6454  &0.0124  &     0.0108   &    0.0111  &     0.0096  &     0.0112  &     0.0109  &     0.0123     \\
2.9500  &0.0108  &     0.0126  &     0.0101   &    0.0096  &     0.0111  &     0.0122  &     0.0116
\end{tabular}
  \end{center}
\caption{DATA: $\mathcal{E}_{MC}$ @CatA}
\end{table}

\begin{table}
\tiny
   \tabcolsep=0.15cm
\renewcommand{\arraystretch}{1.5}
  \begin{center}
\begin{tabular}{c | ccccccc}
$\sqrt{s}$ & BIN1 & BIN2 & BIN3 &  BIN4 &  BIN5 &  BIN6 &  BIN7   \\
\hline
\hline
2.0000  & 0.0066  &      0.0062  &      0.0064  &      0.0061  &      0.0058   &     0.0060   &     0.0069    \\
2.0500  & 0.0095  &      0.0142  &      0.0136  &      0.0139  &      0.0134  &      0.0141   &     0.0099   \\
2.1250  & 0.0187   &     0.0269  &      0.0278  &      0.0280  &      0.0277   &     0.0264  &      0.0192   \\
2.1500  & 0.0226   &     0.0282   &     0.0327  &      0.0306  &      0.0321  &      0.0312  &      0.0233   \\
2.1750  & 0.0250   &     0.0323   &     0.0342  &      0.0346  &      0.0343  &      0.0323  &      0.0231   \\
2.2000  & 0.0284  &      0.0364   &     0.0374  &      0.0374  &      0.0371  &      0.0365  &      0.0293   \\
2.2324  & 0.0283   &     0.0372  &      0.0381  &      0.0396  &      0.0383  &      0.0373  &      0.029   \\
2.3094  & 0.0349   &     0.0397  &      0.0408  &      0.0412  &      0.0417  &     0.0419   &     0.0324   \\
2.3864  & 0.0420  &      0.0515  &      0.0511  &      0.0496  &      0.0535  &     0.0513   &      0.0435   \\
2.3960  & 0.0426   &     0.0489  &      0.0503   &     0.0500  &      0.0523  &      0.0506  &      0.0417   \\
2.6454  & 0.0427  &      0.0530  &      0.0527  &      0.0518  &      0.0554  &      0.0520  &      0.0425  \\
2.9500  & 0.0424   &     0.0533  &      0.0577   &     0.0523  &        0.0548  &      0.0557  &      0.0428
\end{tabular}
  \end{center}
\caption{DATA: $\mathcal{E}_{MC}$ @CatB}
\end{table}

\begin{table}
\tiny
   \tabcolsep=0.15cm
\renewcommand{\arraystretch}{1.5}
  \begin{center}
\begin{tabular}{c | ccccccc}
$\sqrt{s}$ & BIN1 & BIN2 & BIN3 &  BIN4 &  BIN5 &  BIN6 &  BIN7   \\
\hline
\hline
2.0000  & 0.0051  &      0.0072  &      0.0077  &      0.0067  &      0.0071   &     0.0079  &      0.0058      \\
2.0500  & 0.0130  &      0.0161  &      0.0162  &      0.0149   &     0.0166  &      0.0167   &     0.0138    \\
2.1250  & 0.0233   &     0.0289  &      0.0315  &      0.0274   &     0.0309  &      0.0299  &      0.0246    \\
2.1500  & 0.0258  &      0.0326  &      0.0337  &      0.0323  &      0.0343  &      0.0359   &     0.0295    \\
2.1750  & 0.0293  &      0.0366   &     0.0371  &      0.0322  &      0.0377  &      0.0331  &      0.0285    \\
2.2000  & 0.0301   &     0.0383  &      0.0364  &      0.0406   &     0.0396  &      0.0406  &      0.0332    \\
2.2324  & 0.0356  &      0.0419  &      0.0434  &      0.0384   &     0.0446  &      0.0428  &      0.0359    \\
2.3094  & 0.0381   &     0.0429  &      0.0441  &      0.0405  &      0.0434   &     0.0430  &      0.0372    \\
2.3864  & 0.0464   &     0.0523  &      0.0510  &      0.0496  &      0.0529  &      0.0564   &     0.0480 \\
2.3960  & 0.0432   &     0.0527  &      0.0508   &     0.0498   &     0.0515   &     0.0536   &     0.0475    \\
2.6454  & 0.0567   &     0.0671  &      0.0659  &      0.0592  &      0.0639   &     0.0661  &      0.0568    \\
2.9500  & 0.0489  &      0.0604  &      0.0605   &     0.0561  &      0.0600  &      0.0633  &      0.0511   \\
\end{tabular}
  \end{center}
\caption{DATA: $\mathcal{E}_{MC}$ @CatC}
\end{table}

\begin{table}
\tiny
   \tabcolsep=0.15cm
\renewcommand{\arraystretch}{1.5}
  \begin{center}
\begin{tabular}{c | cccccccccccccc}
$\sqrt{s}$ & BIN1 & Err1 & BIN2 & Err2 & BIN3 & Err3 &  BIN4 & Err4 &  BIN5 & Err5 &  BIN6 & Err6 &  BIN7 & Err7  \\
\hline
\hline
2.0000  & 1.23  &   0.10   &  1.04    & 0.09   &  1.29   &  0.12  &   0.95  &   0.08    & 1.20   &  0.10  &   1.11  &   0.09   &  1.13   &  0.09    \\
2.0500  & 1.00  &   0.07  &   1.10  &   0.07   &  1.17   &  0.09   &  0.97    & 0.07 &    0.97  &   0.07   &  1.11   &  0.08   &  1.27   &  0.08  \\
2.1250  & 1.01   &  0.06 &    0.99  &   0.06   &  0.95  &   0.06  &   1.00  &   0.06  &   0.97  &   0.06  &   1.02  &   0.06   &  1.03  &   0.06  \\
2.1500  & 0.95  &   0.05  &   1.05  &   0.06  &   0.92   &  0.05   &  0.88   &  0.05  &   0.92   &  0.06  &   0.99   &  0.06  &   1.02   &  0.06  \\
2.1750  & 0.94   &  0.05   &  0.99 &    0.06  &   1.03   &  0.06   &  0.97  &   0.06  &   0.90  &   0.05  &   1.02   &  0.06   &  0.94   &  0.05  \\
2.2000  & 0.98   &  0.05  &   0.93   &  0.05  &   0.93   &  0.05   &  0.98  &   0.06  &   1.08   &  0.06  &   0.97   &  0.05  &   0.94   &  0.05  \\
2.2324  & 1.01   &  0.05  &   1.03   &  0.06   &  0.99  &   0.05   &  0.93   &  0.05   &  1.00   &  0.06  &   0.95  &   0.05   &  1.09  &   0.06  \\
2.3904  & 0.92  &   0.05  &   0.98   &  0.05   &  0.96  &   0.05   &  0.95  &   0.05   &  0.99  &   0.05   &  0.98  &   0.05  &   1.04   &  0.05   \\
2.3864  & 0.98   &  0.05   &  0.91   &  0.04  &   1.03  &   0.05   &  0.97  &   0.05  &   0.87  &   0.04  &   0.94  &   0.05 &    0.98   &  0.05  \\
2.3960  & 0.94  &   0.04  &   0.89   &  0.04   &  1.01  &   0.05   &  0.94   &  0.04  &   0.90   &  0.04   &  0.95  &   0.04   &  0.96  &   0.04  \\
2.6454  & 0.91  &   0.04  &   0.80  &   0.03  &   0.89 &    0.03  &   0.74  &   0.03   &  0.85  &   0.03  &   0.84  &   0.03  &   0.85   &  0.04  \\
2.9500  & 0.80  &   0.03  &   0.81  &   0.03   &  0.75  &   0.03  &   0.73  &   0.03  &   0.77   &  0.03  &   0.78   &  0.03  &   0.86   &  0.04
\end{tabular}
  \end{center}
\caption{DATA: $\mathcal{C}_{dm}$ @CatA}
\end{table}

\begin{table}
\tiny
   \tabcolsep=0.15cm
\renewcommand{\arraystretch}{1.5}
  \begin{center}
\begin{tabular}{c | cccccccccccccc}
$\sqrt{s}$ & BIN1 & Err1 & BIN2 & Err2 & BIN3 & Err3 &  BIN4 & Err4 &  BIN5 & Err5 &  BIN6 & Err6 &  BIN7 & Err7  \\
\hline
\hline
2.0000  & 0.92   &  0.01   &  0.82  &   0.01   &  0.79 &    0.01   &  0.74   &  0.01   &  0.72  &   0.01  &   0.73   &  0.01   &  0.76   &  0.01    \\
2.0500  & 0.67  &   0.01  &   0.69  &   0.01  &   0.68  &   0.01   &  0.67   &  0.01   &  0.67  &   0.01  &   0.68  &   0.01  &   0.69   &  0.01  \\
2.1250  & 0.79  &   0.01  &   0.78   &  0.01   &  0.76  &   0.01   &  0.76  &   0.01   &  0.75   &  0.01  &   0.77  &   0.01  &   0.78   &  0.01  \\
2.1500  & 0.84  &   0.04   &  0.82  &   0.05  &   0.80   &  0.05  &   0.79  &   0.05   &  0.78  &   0.05  &   0.79  &   0.05   &  0.80  &   0.04  \\
2.1750  & 0.86  &   0.04  &   0.83  &   0.05   &  0.81  &   0.05  &   0.80   &  0.05   &  0.80  &   0.05   &  0.80  &   0.05  &   0.81   &  0.04  \\
2.2000  & 0.83   &  0.04   &  0.81  &   0.05  &   0.81  &   0.05  &   0.80  &   0.05   &  0.80  &   0.04  &   0.81   &  0.05  &   0.81   &  0.04  \\
2.2324  & 0.86   &  0.04  &   0.84  &   0.05  &   0.83   &  0.05   &  0.82   &  0.05  &   0.82    & 0.05  &   0.82   &  0.05  &   0.83   &  0.04  \\
2.3094  & 0.88  &   0.04  &   0.86   &  0.04   &  0.85  &   0.04   &  0.85  &   0.04   &  0.84   &  0.04   &  0.84   &  0.04   &  0.86   &  0.04  \\
2.3864  & 0.88  &   0.01  &   0.89 &    0.01  &   0.88   &  0.01   &  0.88  &   0.01   &  0.88  &   0.01  &   0.88  &   0.01  &   0.89   &  0.01  \\
2.3960  & 0.88  &   0.01   &  0.89  &   0.01  &   0.88  &   0.01   &  0.88  &   0.01   &  0.88  &   0.01  &   0.88  &   0.01  &   0.89   &  0.01  \\
2.6454  & 0.86   &  0.01   &  0.84  &   0.01   &  0.84   &  0.01   &  0.83   &  0.01   &  0.83  &   0.01  &   0.83   &  0.01   &  0.84   &  0.01  \\
2.9500  & 0.80  &   0.01   &  0.79  &   0.01   &  0.80  &   0.01  &   0.80  &   0.01  &   0.80  &   0.01  &   0.80   &  0.01   &  0.81  &   0.01
\end{tabular}
  \end{center}
\caption{DATA: $\mathcal{C}_{dm}$ @CatB}
\end{table}

\begin{table}
\tiny
   \tabcolsep=0.15cm
\renewcommand{\arraystretch}{1.5}
  \begin{center}
\begin{tabular}{c | cccccccccccccc}
$\sqrt{s}$ & BIN1 & Err1 & BIN2 & Err2 & BIN3 & Err3 &  BIN4 & Err4 &  BIN5 & Err5 &  BIN6 & Err6 &  BIN7 & Err7  \\
\hline
\hline
2.0000  & 1.23   &  0.13   &  1.33   &  0.12 &    1.38   &  0.14   &  1.22  &   0.12  &   1.29   &  0.13   &  1.32   &  0.12   &  1.78   &  0.18  \\
2.0500  & 1.21  &   0.07   &  1.29  &   0.07  &   1.27  &   0.07   &  1.19   &  0.07  &   1.28   &  0.07  &   1.22   &  0.07  &   1.34   &  0.08\\
2.1250  & 1.24  &   0.06   &  1.28  &   0.06   &  1.22  &   0.06   &  1.17  &   0.06   &  1.24   &  0.06  &   1.20  &   0.06  &   1.31  &   0.07\\
2.1500  & 1.30   &  0.07  &   1.24 &    0.06   &  1.18  &   0.06  &   1.16   &  0.06  &   1.16   &  0.06  &   1.19   &  0.06   &  1.35   &  0.07\\
2.1750  & 1.31  &   0.08   &  1.24  &   0.06  &   1.17  &   0.06   &  1.16   &  0.06   &  1.15  &   0.06  &   1.19  &   0.06  &   1.35   &  0.08\\
2.2000  & 1.31   &  0.08  &   1.23  &   0.06  &   1.17  &   0.06   &  1.16  &   0.06   &  1.15 &    0.06   &  1.19  &   0.06  &   1.35   &  0.08\\
2.2324  & 1.29 &    0.05 &    1.20 &    0.04  &   1.17  &   0.04  &   1.13  &   0.04  &   1.13  &   0.04  &   1.21  &   0.04   &  1.33   &  0.05\\
2.3094  & 1.27   &  0.07   &  1.21  &   0.06  &   1.18  &   0.06   &  1.11  &   0.06 &    1.13  &   0.06  &   1.28  &   0.06   &  1.34   &  0.07\\
2.3864  & 1.14  &   0.06   &  1.15  &   0.05  &   1.13  &   0.05   &  1.13  &   0.05 &    1.11   &  0.05   &  1.18  &   0.05  &   1.24   &  0.06\\
2.3960  & 1.15   &  0.05  &   1.15   &  0.05  &   1.13  &   0.05   &  1.13   &  0.05  &   1.12  &   0.05   &  1.18   &  0.05 &    1.24   &  0.06\\
2.6454  & 0.95   &  0.03  &   1.03   &  0.03   &  1.07  &   0.03   &  1.03   &  0.03  &   1.0   &  0.03  &   0.96   &  0.03   &  1.01 &    0.03\\
2.9500 &  1.09  &   0.06  &   1.06  &   0.05  &   1.08  &   0.05   &  1.01  &   0.05  &   1.0  &   0.05  &   1.05   &  0.05  &   1.09 &    0.06
\end{tabular}
  \end{center}
\caption{DATA: $\mathcal{C}_{dm}$ @CatC}
\end{table}

\begin{table}
\tiny
   \tabcolsep=0.15cm
\renewcommand{\arraystretch}{1.5}
  \begin{center}
\begin{tabular}{c | cccccccccccccc}
$\sqrt{s}$ & BIN1 & Err1 & BIN2 & Err2 & BIN3 & Err3 &  BIN4 & Err4 &  BIN5 & Err5 &  BIN6 & Err6 &  BIN7 & Err7  \\
\hline
\hline
2.0000  & 0.983 & 0.007 & 0.981 & 0.007 & 0.982 & 0.007 & 0.978 & 0.007 & 0.982 & 0.007 & 0.985 & 0.007 & 0.984 & 0.007 \\
2.0500  & 1.079 & 0.008 & 1.074 & 0.008 & 1.072 & 0.008 & 1.079 & 0.008 & 1.075 & 0.008 & 1.081 & 0.008 & 1.079 & 0.008 \\
2.1250  & 1.244 & 0.009 & 1.227 & 0.009 & 1.222 & 0.009 & 1.218 & 0.009 & 1.220 & 0.009 & 1.221 & 0.009 & 1.243 & 0.009 \\
2.1500  & 1.279 & 0.010 & 1.270 & 0.010 & 1.261 & 0.010 & 1.249 & 0.010 & 1.255 & 0.010 & 1.267 & 0.010 & 1.288 & 0.010 \\
2.1750  & 1.310 & 0.010 & 1.291 & 0.010 & 1.279 & 0.010 & 1.268 & 0.010 & 1.273 & 0.010 & 1.287 & 0.010 & 1.315 & 0.010 \\
2.2000  & 1.316 & 0.010 & 1.286 & 0.010 & 1.278 & 0.010 & 1.266 & 0.010 & 1.271 & 0.010 & 1.287 & 0.010 & 1.319 & 0.010 \\
2.2324  & 1.287 & 0.010 & 1.248 & 0.010 & 1.247 & 0.010 & 1.223 & 0.010 & 1.231 & 0.010 & 1.256 & 0.010 & 1.281 & 0.010 \\
2.3094  & 1.147 & 0.009 & 1.124 & 0.009 & 1.106 & 0.009 & 1.097 & 0.008 & 1.115 & 0.009 & 1.123 & 0.009 & 1.155 & 0.009 \\
2.3864  & 1.112 & 0.008 & 1.082 & 0.008 & 1.070 & 0.008 & 1.063 & 0.008 & 1.071 & 0.008 & 1.085 & 0.008 & 1.109 & 0.008 \\
2.3960  & 1.118 & 0.009 & 1.093 & 0.009 & 1.079 & 0.009 & 1.078 & 0.009 & 1.082 & 0.009 & 1.095 & 0.009 & 1.122 & 0.009 \\
2.6454  & 1.532 & 0.013 & 1.439 & 0.013 & 1.380 & 0.012 & 1.374 & 0.012 & 1.369 & 0.012 & 1.434 & 0.012 & 1.526 & 0.013 \\
2.9500  & 2.114 & 0.021 & 1.845 & 0.019 & 1.784 & 0.018 & 1.698 & 0.018 & 1.701 & 0.018 & 1.842 & 0.019 & 2.132 & 0.022
\end{tabular}
  \end{center}
\caption{DATA: $(1+\delta )$ @all categories}
\end{table}

\begin{table}
\tiny
   \tabcolsep=0.15cm
\renewcommand{\arraystretch}{1.5}
  \begin{center}
\begin{tabular}{c | cc|cc|cc|cc}
$\sqrt{s}$ & $\mathcal{C}_{trg}^{A}$ & $\delta\mathcal{C}_{trg}^{A}$ & $\mathcal{C}_{trg}^{B}$ & $\delta\mathcal{C}_{trg}^{B}$ & $\mathcal{C}_{trg}^{C}$ & $\delta\mathcal{C}_{trg}^{C}$ & $\mathcal{C}_{ex}^{C}$ & $\delta\mathcal{C}_{ex}^{C}$\\
\hline
\hline
2.0000  & 0.78  &  0.01 & 0.87 &  0.03 & 0.89 &  0.03 & 0.75 &  0.02 \\
2.0500  & 0.78  &  0.01 & 0.87 &  0.02 & 0.91 &  0.02 & 0.75 &  0.02 \\
2.1250  & 0.80  &  0.01 & 0.89 &  0.01 & 0.93 &  0.01 & 0.80 &  0.01 \\
2.1500  & 0.81  &  0.01 & 0.90 &  0.01 & 0.93 &  0.01 & 0.83 &  0.01 \\
2.1750  & 0.79  &  0.01 & 0.84 &  0.01 & 0.92 &  0.01 & 0.86 &  0.02 \\
2.2000  & 0.80  &  0.01 & 0.85 &  0.01 & 0.92 &  0.01 & 0.88 &  0.02 \\
2.2324  & 0.81  &  0.01 & 0.85 &  0.01 & 0.93 &  0.01 & 0.89 &  0.02 \\
2.3904  & 0.83  &  0.01 & 0.88 &  0.01 & 0.95 &  0.01 & 0.90 &  0.02 \\
2.3864  & 0.85  &  0.01 & 0.95 &  0.01 & 0.96 &  0.01 & 0.90 &  0.02 \\
2.3960  & 0.85  &  0.01 & 0.94 &  0.01 & 0.96 &  0.01 & 0.90 &  0.02 \\
2.6454  & 0.93  &  0.02 & 0.96 &  0.01 & 0.97 &  0.01 & 0.90 &  0.03 \\
2.9500  & 0.95  &  0.02 & 0.98 &  0.01 & 0.98 &  0.01 & 0.90 &  0.05
\end{tabular}
  \end{center}
\caption{DATA: $\mathcal{C}_{trg}$ @all categories AND $\mathcal{C}_{ex}$ @CatC}
\end{table}

\end{appendices}

\end{document}